\documentclass{emulateapj}
\usepackage{lscape}

\def\sun{\ifmmode\odot\else$\odot$\fi}
\def\HeI{\hbox{He\,{\sc i}}}
\def\H2{\hbox{H$_{2}$}}

\def\Brgamma{\hbox{\rm Br$\gamma$}}

\def\Msun{M$_{\odot}$}
\def\Lsun{L$_{\odot}$}
\def\mic{$\mu$m}

\shorttitle{Resolving the Stellar Populations in NGC~7469}
\shortauthors{T. D\'{\i}az-Santos et al.}

\begin{document}

\title{Resolving the Stellar Populations in the Circumnuclear Ring of NGC~7469}


\author{T.~D\'{\i}az-Santos\altaffilmark{1},
        A.~Alonso-Herrero\altaffilmark{1}, L.~Colina\altaffilmark{1},
        S.~D.~Ryder\altaffilmark{2}, and J.~H.~Knapen\altaffilmark{3}}
\altaffiltext{1}{Departamento de Astrof\'{\i}sica Molecular e Infrarroja, Instituto de Estructura de la Materia, CSIC, E-28006 Madrid, Spain}
\altaffiltext{2}{Anglo-Australian Observatory, P.O. Box 296, Epping, NSW 1710, Australia}
\altaffiltext{3}{Centre for Astrophysics Research, University of Hertfordshire, Hatfield, Herts AL10 9AB, UK}
%\altaffiltext{4}{Gemini Observatory, 670 N. A'Ohoku Place, Hilo, HI 96720, U.S.A.}

\begin{abstract}

We investigate the stellar populations in the star forming ring of the 
luminous infrared galaxy NGC~7469.  We use
{\it Hubble Space Telescope}
multi-wavelength (UV through NIR) imaging complemented
with new $K$-band ground-based long-slit spectroscopy, and
mid-IR and radio maps from the literature.  
Spectral energy distributions (SEDs) and evolutionary synthesis models
have been used to characterize the star formation at different scales
from those
of individual star clusters (tens of pc) to that of the entire star-forming
ring (kpc scale). At the smallest scales two different populations 
of massive
($1-10 \times 10^6\,M_\odot$) clusters are identified. About
25\% of the clusters are young
($1-3\,$Myr) and extincted ($A_V$ $\approx$ 3 mag), whereas the vast
majority are
of intermediate age ($\sim 9$ to $20\,$Myr) and less obscured
($A_V$ $\approx$ 1 mag).
At larger (hundreds of pc) scale, an analysis of the
integrated SED and spectroscopic data of the ring
indicates the presence of two stellar populations. 
The young ($5-6\,$Myr) and obscured stellar 
population accounts for the \Brgamma\, emission and most of the
IR luminosity, and about one-third of the stellar mass
of the ring.
The much less obscured intermediate-age
population has properties similar to those of the
majority of the (older) $1.1\,\mu$m-selected star
clusters. The
distribution of these two populations is clearly different and even spatially
anti-correlated. The UV-optical-NIR continuum (including the
majority of the clusters) of the ring
traces mostly the mildly obscured
intermediate-age population, while the MIR and radio peaks mark the location of
the youngest and obscured star-forming regions. Moreover, the two
brightest MIR and radio peaks are spatially coincident with the ends of
the nuclear molecular gas bar.
%and probably are located near the Inner Lindblad Resonance of the large-scale stellar bar.
% where most of the star formation is expected to take place, thus reinforcing the general picture in which the star formation in the ring is fuelled by gaseous inflow induced by the bar. 
This study emphasizes the need for multi-wavelength, high-angular resolution
observations to characterize the star formation in the dust-obscured regions
commonly present in LIRGs.

\end{abstract}

\keywords{galaxies: individual (NGC~7469) --- galaxies: clusters: individual --- galaxies: star clusters --- galaxies: starburst --- galaxies: stellar content --- infrared: galaxies}

%________________________________________________________________
\section[]{Introduction}\label{s:intro}

Star formation (SF) in Luminous Infrared Galaxies (LIRGs, $10^{11}\, \leq$
$L_{\rm IR [8-1000\mu m]}\, \leq\, 10^{12}\,$\Lsun,  see
Sanders \& Mirabel
1996, and Lonsdale, Farrah \& Smith 2006 for detailed reviews) is one of
the most energetic phenomena in the Universe.
In a large fraction of local LIRGs the
bulk of their SF is generated
within their innermost regions (a few kpc), in ring-like structures,
mini-spiral arms, and regions of compact emission
%Lipari et al. 2000;
(Soifer et al. 2001;
%Murphy et al. 2001;
%Elmegreen et al. 2002; Smith et al. 2005;
Alonso-Herrero et al. 2002, 2006a).
These regions of SF
contain large amounts of molecular gas 
($\gtrsim 10^{8}\,$\Msun, Scoville et al. 1991
Downes \& Solomon 1998; Gao \& Solomon 2004)
and dusty, young ($\lesssim 100\,$Myr),
massive ($10^{5-7}\,$\Msun) super star clusters (SSCs;
%Ravindranath \& Prabhu 1998; Smith et al. 1999; %Reunanen et al. 2000; 
see, among others, Surace, Sanders, \& Evans 2000; Scoville et al. 2000;
Alonso-Herrero et al. 2000,
%2001;
2002; Wilson et al. 2006). 
%Elmegreen et al. 2002;
%de Grijs et al. 2003; Smith et al. 2005, among many others). 
The intense SF is often related to past interactions
suffered by the host galaxy
with a close companion and/or to the presence of a bar
(see Knapen 2004 for a review). Both mechanisms are
believed to be efficient in driving large amounts of gas towards the nuclear
regions.

NGC~7469 is a SABa galaxy with a Seyfert 1 nucleus, 
located at a distance of $\sim$~65~Mpc ($cz$~$\simeq 4917\,$km~s$^{-1}$;
Heckman et al. 1986; H$_0\, =\, 75\,$km~s$^{-1}$). A companion
galaxy, IC~5283, located at $\sim\, 22\,$kpc (Burbidge,
Burbidge \& Prendergast 1963;
Dahari 1985), is believed to have interacted with NGC~7469
more than $1.5\, \times\, 10^8\,$yr
ago (Genzel et al. 1995, hereafter G95). NGC~7469 harbors a circumnuclear
star-forming ring with an approximate diameter of  $5\arcsec$
($\sim\, 1.6\,$kpc) which accounts for
approximately two-thirds of the bolometric luminosity of the galaxy.
This circumnuclear emission  
was first detected in the radio and in the near-infrared (NIR)
(Ulvestad, Wilson,
\& Sramek 1981, and Cutri et al. 1984, respectively). The ring was later
observed in the optical (Wilson et al. 1986;
Mauder et al. 1994; Malkan, Gorjian, \& Tam 1998), NIR (G95),
%; Davies et al. 2004)
mid-infrared (MIR, Miles et al. 1994; Soifer et al. 2003; Gorjian et al. 2004),
far-infrared (FIR, Papadopoulos et al. 2000), and again at radio wavelengths
(Meixner et al. 1990;
Wilson et al. 1991). NGC~7469 has a large-scale (several kpc) stellar bar
detected in the NIR (Knapen, Shlosman \& Peletier 2000) that gives the
galaxy its 'SAB' classification,
but also harbors a nuclear gas bar (with a size similar to the diameter
of the ring) that has been reported by various authors (Laine et al. 2002;
Davies, Tacconi, \& Genzel 2004). A bright radio supernova (SN2000ft) has been
discovered in the circumnuclear ring (Colina et al. 2001; Alberdi et al.
2006), and recently its optical counterpart has been identified
(Colina et al. 2007).

G95 presented a combination of NIR
broad-band speckle and Fabry-Perot imaging, as well as
integral-field spectroscopy of the nuclear regions  of NGC~7469 at
sub-arcsecond resolution. They were able to separate the emission from
the unresolved Seyfert nucleus from that of the surrounding
starburst ring. By comparing multi-wavelength diagnostics for the ring
as a whole with their starburst model, they concluded that SF 
in the ring has either: (a) been progressing at a constant
rate for the past several $10^7\,$yr, but with few high-mass stars;
or (b) been
decaying exponentially since the onset of a burst 15~Myr ago, with a
fairly `normal' initial mass function (IMF). However, they were unable
to assess the possible variation of the SF history {\em along\/} the ring.

In this paper we present \textit{Hubble Space Telescope (HST)}
UV through NIR imaging, as well as low resolution ($R\sim600$) $K$-band
spectroscopy at four different
locations along the circumnuclear star-forming ring of NGC~7469.
Our goal is to study in
detail the SF properties of individual star clusters (spatial scales
of tens of parsecs) 
detected in the ring (see, e.g., Scoville et al. 2000), of ring sections
(spatial scales of a few hundred parsecs), as well as of the ring as a whole. 
This work is organized as follows: in \S~\ref{s:obs} we present
imaging and spectroscopic data and their reduction procedures and analysis;
%in \S~\ref{s:morph} we describe the morphology of the SF ring;
in \S~\ref{s:model} we
describe the fitting method employed to obtain the parameters of the
stellar populations;
in \S~\ref{s:results_indiv} we present the results for the smallest scales
(individual clusters) whereas results for the ring sections and the
whole ring are given in \S~\ref{s:results_whole}; finally, in \S~\ref{s:discu}
we discuss the results and in \S~\ref{s:conclu} briefly summarize
the main conclusions.

\section[]{Observations and Data Analysis}\label{s:obs}

\subsection[]{{\it HST}  Archival Imaging}\label{s:images}

\begin{figure*}
\epsscale{.9}
\plotone{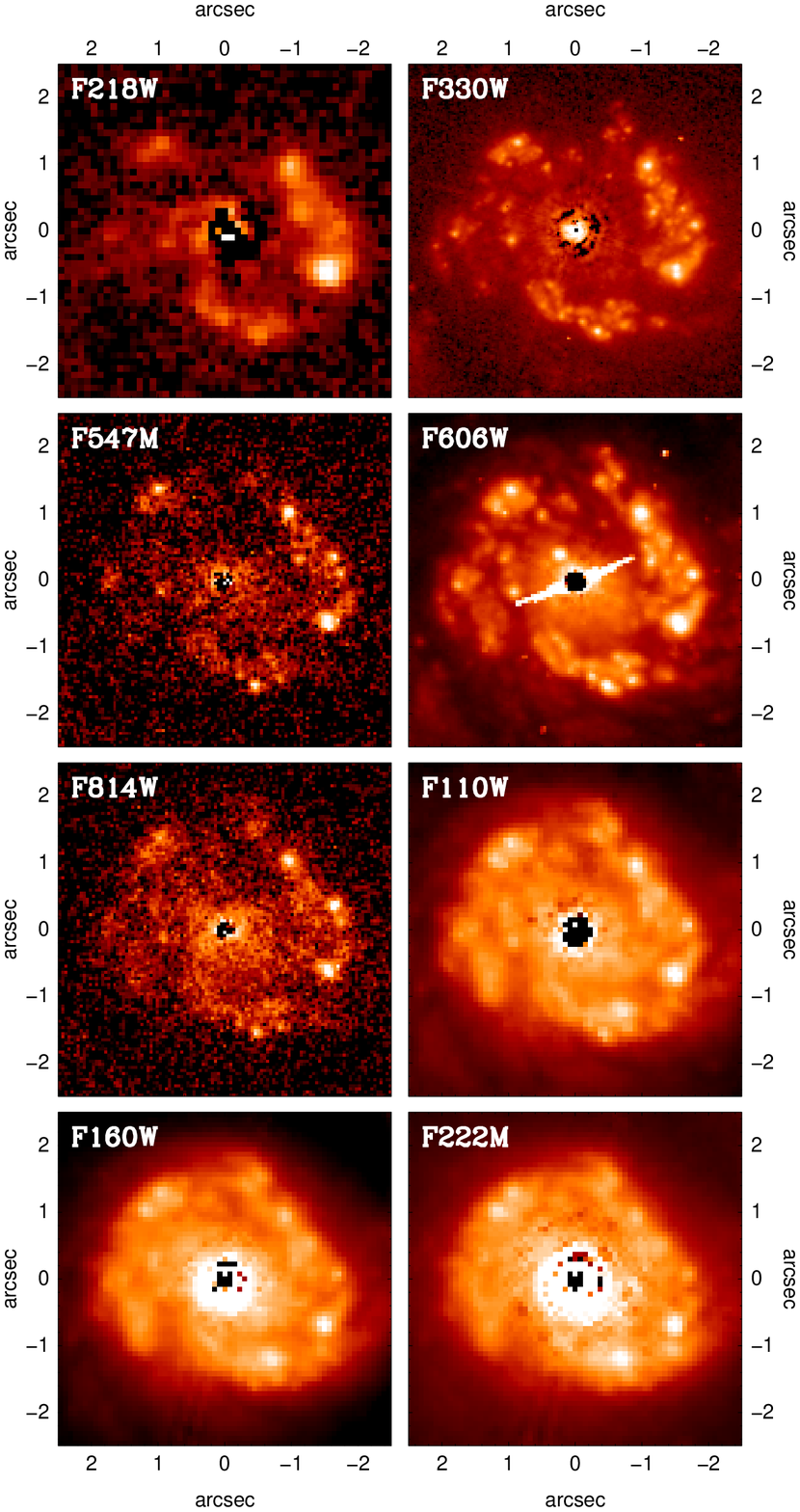}
\caption{\footnotesize Images of the central $5\, \arcsec$ ($\sim 1.6\,$kpc) of NGC~7469 showing the ring of SF in all the available {\it HST} filters. The images are shown in a logarithmic scale. Note that we have attempted to subtract the nuclear source using PSFs generated with TinyTim. The orientation is north up, east to the left. [\textit{See the electronic edition of the Journal for a color version of this figure.}]}\label{f:panel}
\end{figure*}

We have retrieved {\it HST}  archival images of the central region of
NGC~7469 taken through eight filters using three different instruments:
WFPC2, NICMOS, and ACS. The images cover a broad wavelength range  going
from the UV (0.218~$\mu$m) to the NIR (2.22~$\mu$m). Detailed information
about the \textit{HST} observations is given in Table~\ref{t:filters}.
The UV and optical images were calibrated employing the \textit{On The Fly
Recalibration} (OTFR) system. We reduced and calibrated the NICMOS
images as described by 
Alonso-Herrero et al. (2000). We subtracted theoretical PSFs
generated with the TinyTim software
\footnote{\texttt{http://www.stsci.edu/software/tinytim}} (Krist et al. 1998)
for each filter in an attempt to minimize the Seyfert nucleus 
contribution when analyzing the properties of the ring of SF. 
The {\it HST} images of the central $\simeq 1.6\,$kpc of NGC~7469 are
presented in Fig.~\ref{f:panel}.

\begin{deluxetable*}{lccccccccc}
\tabletypesize{\scriptsize}
%\rotate
%\tablewidth{0pc}
%\tablenum{}
%\tablecolumns{9}
\tablecaption{\scriptsize HST Imaging data}
\tablehead{\colhead{Filter} & \colhead{Instrument} & \colhead{Detector} & \colhead{Plate scale} & \colhead{Date} & \colhead{t$_{\rm exp}$\tablenotemark{(a)}} & \colhead{Orientation} & \colhead{TT FWHM\tablenotemark{(b)}} & \colhead{TT FWHM\tablenotemark{(b)}} & \colhead{ID : PI\tablenotemark{(c)}} \\
\cline{8-9}
\colhead{} & \colhead{} & \colhead{} & \colhead{(arcsec pixel$^{-1}$)} & \colhead{(yy/mm/dd)} & \colhead{(s)} & \colhead{(degrees)} & \colhead{(arcsec)} & \colhead{(pc)} & \colhead{}}
\startdata
F218W & WFPC2  &  WF3 & 0.1    & 99/06/28 & 1200 &   107.54 & 0.176 & 55 & 6358 : Colina \\
F330W & ACS    &  HRC & 0.026  & 02/11/20 & 1140 & --107.75 & 0.051 & 16 & 9379 : Schmitt\\
F547M & WFPC2  &   PC & 0.0455 & 00/05/13 &   12 &  --61.36 & 0.076 & 24 & 8240 : Wilson \\
F606W & WFPC2  &   PC & 0.0455 & 94/06/10 &  500 &  --68.35 & 0.073 & 23 & 5479 : Malkan \\
F814W & WFPC2  &   PC & 0.0455 & 00/05/13 &    7 &  --61.36 & 0.085 & 27 & 8240 : Wilson \\
F110W & NICMOS & NIC2 & 0.0755 & 97/11/10 &  352 & --150.15 & 0.117 & 37 & 7219 : Scoville \\
F160W & NICMOS & NIC2 & 0.0755 & 97/11/10 &  352 & --150.15 & 0.154 & 48 & 7219 : Scoville \\
F222M & NICMOS & NIC2 & 0.0755 & 97/11/10 &  480 & --150.15 & 0.202 & 64 & 7219 : Scoville \\
\enddata
\tablecomments{\scriptsize (a) Total exposure time of the final image; (b) FWHM of TinyTim PSFs; (c) Proposal ID and PI.}\label{t:filters}
\end{deluxetable*}

\subsubsection{Star Cluster selection}\label{c:clusteranalysis}

Star clusters were selected at $1.1\,\mu$m (NICMOS F110W filter) because of 
two main reasons: (a) the NIR spectral region is 
less affected by dust extinction than
UV and optical wavelengths: $A_{UV\, (0.22\mu m)}\, \sim\, 3\, \times\, A_V\, \sim\, 20\, \times\, A_{1.1 \mu m}\,$ and $A_{B (0.33\mu m)}\, \sim\, 2\, \times\, A_V\, \sim\, 13\, \times\, A_{1.1\mu m}\,$(Calzetti 1997, CAL hereafter); (b) the F110W filter has the best spatial resolution 
among the NICMOS filters (see Table~\ref{t:filters}). 

\begin{figure}
\epsscale{1.2}
\plotone{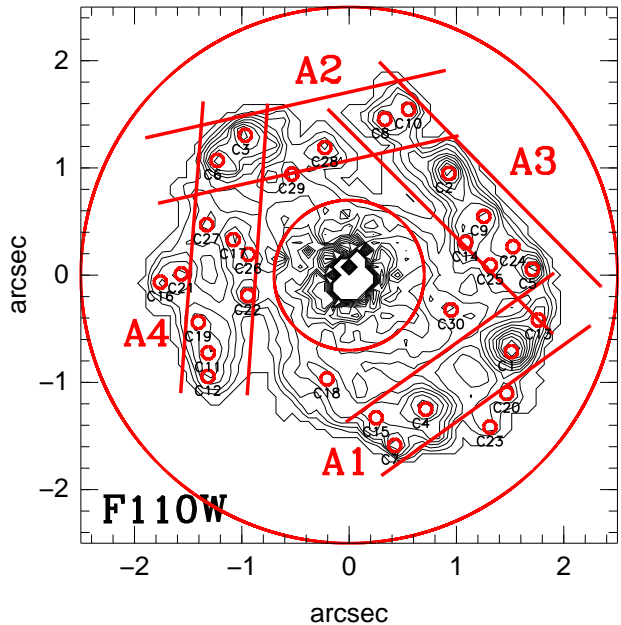}
\caption{\footnotesize The contours, shown in a logarithmic scale, represent the {\it HST}/NICMOS F110W emission. The 1.1~$\mu$m-selected star clusters are marked with small open circles with sizes equivalent to the FWHM of the F110W PSF ($\simeq 37\,$pc, see Table~\ref{t:filters}). The location and P.A. of the UKIRT/CGS4 slits are also marked, wit  the 0$\farcs$6 slit width and extraction lengths (see Table~\ref{t:slitpars}) shown to scale. The ring area defined by G95 (enclosed by two open circles of radii 0$\farcs$7 and 2$\farcs$5, see text) is also shown. North is up, and east to the left. [\textit{See the electronic edition of the Journal for a color version of this figure.}]}\label{f:knotpos}
\end{figure}

The positions of the star clusters in all the {\it HST} images
were determined using a procedure similar to the growing
region method employed by the {\sc HiiPhot} software (Thilker, Braun, 
\& Walterbos 2000). This search method is also similar to that
used by {\sc Sextractor} (Bertin \& Arnouts 1996). A 2$\sigma$ threshold
above a background selected in four regions in the outskirts of the ring
was imposed for cluster identification. However, this background was in some
cases not representative of the strong, complex and
variable ``background'' emission surrounding the clusters 
(that is, underlying unresolved emission within the 
ring acting as a pedestal background).
The search method had to be stopped manually when noise peaks
began to be identified as positive detections in the images, specially
in those with the lowest S/N ratio (F547M and F814W).
A total of 30 clusters in the ring of SF 
was selected at $1.1\, \mu$m. Their locations are given in
Table~\ref{t:knotphot} (see also Fig.~\ref{f:knotpos}). 
Out of the 30 $1.1\,\mu$m star clusters within the ring selected
by our method, only five (C1, C3, C5, C6, and C7) were
identified by Scoville et al. (2000) using the same NIR dataset.

\begin{deluxetable*}{lccrrrrrrrr}
\tabletypesize{\scriptsize}
%\rotate
%\tablewidth{0pc}
%\tablenum{}
%\tablecolumns{11}
\tablecaption{\scriptsize Positions and flux densities of the $1.1\,\mu$m-selected star clusters in the ring of SF}
\tablehead{\colhead{\#\tablenotemark{(a)}} & \colhead{$X$\tablenotemark{(b)}} & \colhead{$Y$\tablenotemark{(b)}} & \colhead{F218W\tablenotemark{(c)}} & \colhead{F330W\tablenotemark{(c)}} & \colhead{F547M\tablenotemark{(c)}} & \colhead{F606W\tablenotemark{(c)}} & \colhead{F814W\tablenotemark{(c)}} & \colhead{F110W\tablenotemark{(c)}} & \colhead{F160W\tablenotemark{(c)}} & \colhead{F222M\tablenotemark{(c)}}}
\startdata
C1 &   -1.54 &  -0.67 &  9.88$\times$10$^{-16}$ &  $_{1\#}$3.44$\times$10$^{-16}$ &  1.03$\times$10$^{-16}$ &  7.73$\times$10$^{-17}$ &  5.90$\times$10$^{-17}$ &  3.58$\times$10$^{-17}$ &  2.79$\times$10$^{-17}$ &  1.93$\times$10$^{-17}$ \\
C2 &   -0.96 &   0.99 &  2.59$\times$10$^{-16}$ &  1.69$\times$10$^{-16}$ &  1.05$\times$10$^{-16}$ &  8.85$\times$10$^{-17}$ &  4.95$\times$10$^{-17}$ &  2.63$\times$10$^{-17}$ &  1.48$\times$10$^{-17}$ &  9.85$\times$10$^{-18}$ \\
C3 &  0.93 &   1.34 &  9.41$\times$10$^{-17}$ &  7.38$\times$10$^{-17}$ &  6.74$\times$10$^{-17}$ &  4.94$\times$10$^{-17}$ &  2.36$\times$10$^{-17}$ &  2.28$\times$10$^{-17}$ &  1.79$\times$10$^{-17}$ &  1.22$\times$10$^{-17}$ \\
C4 &   -0.74 &  -1.20 &       ...  &  6.87$\times$10$^{-18}$ &       ...  &  $_{1\#}$2.96$\times$10$^{-17}$ &       ...  &  1.97$\times$10$^{-17}$ &  1.93$\times$10$^{-17}$ &  1.60$\times$10$^{-17}$ \\
C5 &   -1.73 &   0.10 &  9.23$\times$10$^{-17}$ &  6.42$\times$10$^{-17}$ &  4.03$\times$10$^{-17}$ &  4.67$\times$10$^{-17}$ &  2.27$\times$10$^{-17}$ &  1.80$\times$10$^{-17}$ &  1.24$\times$10$^{-17}$ &  7.93$\times$10$^{-18}$ \\
C6 &  1.19 &   1.11 &  4.95$\times$10$^{-17}$ &  $_{1\#}$2.69$\times$10$^{-17}$ &       ...  &  2.03$\times$10$^{-17}$ &  1.63$\times$10$^{-17}$ &  1.68$\times$10$^{-17}$ &  1.63$\times$10$^{-17}$ &  1.43$\times$10$^{-17}$ \\
C7 &   -0.46 &  -1.54 &  1.29$\times$10$^{-16}$ &  $_{1\#}$1.31$\times$10$^{-16}$ &  7.03$\times$10$^{-17}$ &  7.59$\times$10$^{-17}$ &  3.74$\times$10$^{-17}$ &  1.41$\times$10$^{-17}$ &       ...  &       ...  \\
C8 &   -0.37 &   1.49 &       ...  &  $_{1\#}$8.66$\times$10$^{-18}$ &       ...  &  1.47$\times$10$^{-17}$ &       ...  &  1.34$\times$10$^{-17}$ &  1.02$\times$10$^{-17}$ &  9.85$\times$10$^{-18}$ \\
C9 &   -1.28 &   0.59 &       ...  &  $_{2\#}$3.23$\times$10$^{-17}$ &  1.92$\times$10$^{-17}$ &  $_{1\#}$2.02$\times$10$^{-17}$ &       ...  &  1.22$\times$10$^{-17}$ &  1.18$\times$10$^{-17}$ &  9.39$\times$10$^{-18}$ \\
C10 &   -0.58 &   1.58 &       ...  &  4.83$\times$10$^{-18}$ &       ...  &  1.31$\times$10$^{-17}$ &       ...  &  1.20$\times$10$^{-17}$ &  1.23$\times$10$^{-17}$ &       ...  \\
C11 &  1.27 &  -0.68 &       ...  &  2.89$\times$10$^{-18}$ &       ...  &       ...  &       ...  &  1.09$\times$10$^{-17}$ &  1.06$\times$10$^{-17}$ &  7.67$\times$10$^{-18}$ \\
C12 &  1.27 &  -0.90 &       ...  &  2.89$\times$10$^{-18}$ &       ...  &  8.71$\times$10$^{-18}$ &       ...  &  1.06$\times$10$^{-17}$ &       ...  &       ...  \\
C13 &   -1.79 &  -0.37 &       ...  &  1.36$\times$10$^{-17}$ &       ...  &  1.47$\times$10$^{-17}$ &       ...  &  1.02$\times$10$^{-17}$ &       ...  &  4.35$\times$10$^{-18}$ \\
C14 &   -1.11 &   0.35 &  1.19$\times$10$^{-16}$ &  $_{2\#}$6.62$\times$10$^{-17}$ &  3.16$\times$10$^{-17}$ &  $_{1\#}$3.70$\times$10$^{-17}$ &  1.65$\times$10$^{-17}$ &  1.00$\times$10$^{-17}$ &  6.88$\times$10$^{-18}$ &  6.55$\times$10$^{-18}$ \\
C15 &   -0.28 &  -1.28 &  3.57$\times$10$^{-17}$ &       ...  &  2.14$\times$10$^{-17}$ &  $_{1\#}$3.64$\times$10$^{-17}$ &  1.79$\times$10$^{-17}$ &  8.91$\times$10$^{-18}$ &  8.28$\times$10$^{-18}$ &  7.05$\times$10$^{-18}$ \\
C16 &  1.72 &  -0.02 &  3.67$\times$10$^{-17}$ &  $_{1\#}$3.11$\times$10$^{-17}$ &       ...  &  1.75$\times$10$^{-17}$ &       ...  &  8.89$\times$10$^{-18}$ &  :6.81$\times$10$^{-18}$ &       ...  \\
C17 &  1.04 &   0.37 &       ...  &  1.13$\times$10$^{-17}$ &       ...  &  1.17$\times$10$^{-17}$ &       ...  &  7.82$\times$10$^{-18}$ &  6.99$\times$10$^{-18}$ &       ...  \\
C18 &  0.17 &  -0.92 &  4.22$\times$10$^{-17}$ &  2.03$\times$10$^{-17}$ &       ...  &  2.29$\times$10$^{-17}$ &       ...  &  7.73$\times$10$^{-18}$ &  8.36$\times$10$^{-18}$ &  6.28$\times$10$^{-18}$ \\
C19 &  1.36 &  -0.39 &  1.14$\times$10$^{-17}$ &       ...  &       ...  &       ...  &       ...  &  7.45$\times$10$^{-18}$ &  7.56$\times$10$^{-18}$ &  5.93$\times$10$^{-18}$ \\
C20 &   -1.49 &  -1.05 &       ...  &       ...  &       ...  &       ...  &       ...  &  7.35$\times$10$^{-18}$ &  8.65$\times$10$^{-18}$ &  9.91$\times$10$^{-18}$ \\
C21 &  1.52 &   0.05 &       ...  &  3.06$\times$10$^{-18}$ &       ...  &       ...  &       ...  &  7.20$\times$10$^{-18}$ &  :6.81$\times$10$^{-18}$ &  5.31$\times$10$^{-18}$ \\
C22 &  0.91 &  -0.15 &  4.77$\times$10$^{-17}$ &  3.15$\times$10$^{-17}$ &  3.42$\times$10$^{-17}$ &  2.59$\times$10$^{-17}$ &  1.77$\times$10$^{-17}$ &  6.94$\times$10$^{-18}$ &       ...  &       ...  \\
C23 &   -1.34 &  -1.37 &       ...  &  $_{1\#}$1.18$\times$10$^{-17}$ &       ...  &  7.69$\times$10$^{-18}$ &       ...  &  6.85$\times$10$^{-18}$ &  5.24$\times$10$^{-18}$ &       ...  \\
C24 &   -1.55 &   0.31 &  4.15$\times$10$^{-17}$ &  $_{1\#}$:3.54$\times$10$^{-17}$ &      ...   &  1.60$\times$10$^{-17}$ &        ...  &  6.18$\times$10$^{-18}$ &       ...  &       ...  \\
C25 &   -1.34 &   0.13 &       ...  &  $_{1\#}$:1.37$\times$10$^{-17}$ &       ...  &  7.78$\times$10$^{-18}$ &       ...  &  5.97$\times$10$^{-18}$ &  4.34$\times$10$^{-18}$ &       ...  \\
C26 &  0.90 &   0.24 &       ...  &       ...  &       ...  &       ...  &       ...  &  4.43$\times$10$^{-18}$ &       ...  &       ...  \\
C27 &  1.29 &   0.51 &       ...  &  2.31$\times$10$^{-18}$ &       ...  &       ...  &       ...  &  4.26$\times$10$^{-18}$ &       ...  &       ...  \\
C28 &  0.19 &   1.23 &       ...  &  4.38$\times$10$^{-18}$ &       ...  &  8.31$\times$10$^{-18}$ &       ...  &  4.09$\times$10$^{-18}$ &  2.87$\times$10$^{-18}$ &  3.41$\times$10$^{-18}$ \\
C29 &  0.50 &   0.98 &       ...  &  1.48$\times$10$^{-17}$ &       ...  &  1.39$\times$10$^{-17}$ &       ...  &  2.36$\times$10$^{-18}$ &       ...  &       ...  \\
C30 &   -0.98 &  -0.28 &       ...  &       ...  &       ...  &       ...  &       ...  &  2.18$\times$10$^{-18}$ &       ...  &       ...  \\
\enddata
\tablecomments{\scriptsize (a) Cluster label. The clusters are sorted by their F110W flux; (b) Positions of the clusters relative to the nucleus (in arcsec). A positive sign indicates west and north directions; (c) Flux densities (in erg s$^{-1}$ cm$^{-2}$ \AA$^{-1}$) at each filter. The number with a hash symbol next to a flux density indicates the number of additional counterparts that have been detected aside from the first one. The quoted flux is the sum of all the counterparts. A colon next to a flux density indicates that the same cluster is associated with two different $1.1\,\mu$m star clusters.}\label{t:knotphot}
\end{deluxetable*}

In order to construct spectral energy distributions (SEDs) 
of the $1.1\,\mu$m-selected star
clusters, we cross-correlated their positions with those of clusters detected
at other wavelengths. Clusters located at distances 
$\le\, 1.5\, \times\, {\rm FWHM\, _{F110W}}$ ($\le\, 0\farcs175\,
\simeq 55\, {\rm pc}$) were identified as counterparts
of the $1.1\, \mu$m clusters. Because the F330W and F606W images have both
good S/N and higher
spatial resolution than the F110W image, a few $1.1\,\mu$m
clusters were found to have 
multiple cross-identifications at these wavelengths. 
Table~\ref{t:ccknots} provides information about cluster identification
and cross-correlation.
%gives the number of clusters
%identified in each {\it HST} image, the number of clusters identified as
%counterparts of the $1.1\,\mu$m-selected clusters for each wavelength, 
%and the fraction of star clusters at a given wavelength identified
%by selecting clusters at $1.1\,\mu$m.
For all the images except those with the highest spatial resolutions (F330W
and F606W), most of the clusters identified at a given wavelength are
indeed those also selected at $1.1\,\mu$m, even in the UV. The
fraction of the F330W and F606W clusters cross-identified at $1.1\,\mu$m
is smaller than in the other filters, but consistent with the larger number of
clusters detected in these two filters.

\begin{deluxetable}{lcccc}
\tabletypesize{\scriptsize}
%\rotate
%\tablewidth{0pc}
%\tablenum{}
%\tablecolumns{5}
\tablecaption{\scriptsize Photometric uncertainties, number of clusters and cross-identifications}
\tablehead{\colhead{Filter} & \colhead{Phot} & \colhead{No.} & \colhead{No.} & \colhead{Fraction of} \\
\colhead{} & \colhead{uncert.\tablenotemark{(a)}} & \colhead{clusters\tablenotemark{(b)}} & \colhead{cross-ID\tablenotemark{(c)}} & \colhead{$1.1\,\mu$m selected\tablenotemark{(d)}}}
\startdata
F218W & 10\% & 16 &  13      & 81\%      \\%&  43 \\
F330W &  5\% & 56 &  25+11   & 45+20\%   \\%&  83 \\
F547M & 15\% & 10 &   9      & 90\%      \\%&  30 \\
F606W &  5\% & 42 &  23+4    & 55+10\%   \\%&  77 \\
F814W & 12\% & 11 &  10      & 91\%      \\%&  33 \\
F110W &  8\% & 30 &  30      & 100\%     \\%& 100 \\
F160W &  8\% & 20 &  20      & 100\%     \\%&  67 \\
F222M &  9\% & 18 &  17      &  94\%     \\%&  57 \\
\enddata
\tablecomments{\scriptsize (a) Average estimated uncertainty for the cluster photometry (see text for details); (b) Number of identified clusters; (c) Number of cross-identified $1.1\,\mu$m clusters in each image plus the number of clusters identified as double or triple counterparts for the F330W and F606W filters; (d) Fraction of cross-identified $1.1\,\mu$m clusters with respect to the total number of clusters detected in each image.}\label{t:ccknots}
\end{deluxetable}

\subsubsection{Cluster photometry}

The cluster intensity was modelled with a 2D Gaussian function whose 
width was fixed to that
measured from theoretical PSFs generated with TinyTim (Krist et al. 1998)
for each filter. 
Additionally, our method fits the emission surrounding the source with a
2D plane, which is then interpolated to the region where the cluster is
located. The region used to fit the background was defined as the area
enclosed by a two pixel wide ring with its inner radius containing the
90\% of the TinyTim PSF flux. This is done after rejecting those pixels
that exceed the
fitted background by $2.75 \sigma$ (i.e., pixels in the background
with possible contamination from other clusters).
%plus a 2D plane, the latter used to simulate the background surrounding the cluster.
A $\chi^2$-minimizing
Marquardt method (Bevington \& Robinson 2003)
implemented in IDL was used to derive 
accurate positions and fluxes of the point sources. However,
since the {\it HST} PSFs are not Gaussians, 
an aperture correction (filter-dependent)
was needed. These corrections were obtained by modelling the individual
theoretical PSFs generated with TinyTim in the same way as the star clusters.

In order to estimate the photometric 
uncertainty of this method, we generated a grid of test images
where TinyTim PSFs with random positions and variable
intensities were placed. The number density of point sources was
approximately 0.01 pixel$^{-2}$. PSFs separated by 
$<2-3\,$pixels were common in the test images, similar to the 
the cluster density observed in the ring of NGC~7469. 
We ran simulations varying the background intensity and standard deviation, 
as well as the PSF intensity with respect to that of the background. 
We applied our fitting method to measure the PSF fluxes and compare them 
with the input values. Our point source modelling method was compared with 
two other classical photometric tools: aperture photometry, and the tasks of
{\sc daophot} (Stetson 1987) in {\sc iraf}\footnote{IRAF is distributed
by the National Optical Astronomy Observatories, which are operated by
the Association of Universities for Research in Astronomy, Inc., under
cooperative agreement with the National Science Foundation.}.
We found that our method is capable of recovering fluxes with an
accuracy typically as high as that of {\sc daophot} and in some cases twice
as high (depending on the filter and the parameters of the simulation),
and $\sim 2-4$ times higher than the classical aperture photometry method.
This is probably due to the detailed background modelling performed by our
method. 

These simulations were also used to estimate the 
uncertainties (see Table~\ref{t:ccknots}) 
of the cluster photometry in each image by selecting the simulation 
that best matched the background parameters (standard deviation
and relative intensity of the clusters with respect to the background)
of the ring.
%standard deviation and cluster fluxes with respect to the background. 
The photometry for the $1.1\,\mu$m-selected star clusters is given in 
Table~\ref{t:knotphot}. In the case of multiple cross-identifications, 
the sum of the flux densities of
all cross-identified clusters at a given wavelength was used as the
representative of the cluster selected at 1.1~$\mu$m.

\subsection[]{Long-slit NIR spectroscopy}\label{s:spec}

Long-slit spectroscopy of the central region of NGC~7469 was 
obtained with the common-user NIR spectrograph CGS4 (Mountain et
al. 1990) on UKIRT during 1999~October~5 and 6. 
We employed the $0\farcs61\,$pixel$^{-1}$ plate scale, a 1~pixel wide
slit, and the 40~line~mm$^{-1}$ grating in first order,
which delivers complete spectral coverage from $1.85-2.45\,\mu$m at a
resolving power of 600.
Conditions were
photometric with $0\farcs6$ seeing throughout the first night, but
increasing amounts of cirrus curtailed observations on the second
night. A total of 4~slit position angles and nuclear
offsets were selected (see Fig.~\ref{f:knotpos}). Details of
the observing run are given in Table~\ref{t:slitpars}. After
setting the slit to the desired position angle, a manual search for
the $K$-band nuclear peak was conducted, following which accurate
offsets were applied by use of the UKIRT crosshead to place the slit
right at the desired locations. In addition, a single
slit position passing right through the nucleus was observed.

\begin{deluxetable}{lcccc}
\tabletypesize{\scriptsize}
%\rotate
%\tablewidth{0pc}
%\tablenum{}
%\tablecolumns{5}
\tablecaption{\scriptsize UKIRT/CGS4 slit parameters}
\tablehead{\colhead{Slit} & \colhead{P.A.} & \colhead{$t_{\rm exp}$\tablenotemark{(a)}} & \colhead{Aperture\tablenotemark{(b)}} & \colhead{Aperture\tablenotemark{(b)}} \\
\colhead{} & \colhead{(degree)} & \colhead{(s)} & \colhead{(arcsec)} & \colhead{(pc)}}
\startdata
A1  &  123.9  &  5280 & 0$\farcs$61 $\times$ 2$\farcs$40 & 192 $\times$ 756 \\
A2  &  107.7  &  5520 & 0$\farcs$61 $\times$ 2$\farcs$93 & 192 $\times$ 923 \\
A3  &   42.1  &  6240 & 0$\farcs$61 $\times$ 3$\farcs$00 & 192 $\times$ 945 \\
A4  &  174.7  &  5280 & 0$\farcs$61 $\times$ 2$\farcs$69 & 192 $\times$ 848 \\
Nucleus & ... &   960 & 0$\farcs$61 $\times$ 1$\farcs$00 & 192 $\times$ 315 \\
\enddata
\tablecomments{\scriptsize (a) Exposure time; (b) Dimensions of extraction apertures.}\label{t:slitpars}
\end{deluxetable}

The data reduction was accomplished using a combination of tasks within the
Starlink {\sc cgs4dr} and NOAO {\sc iraf}
packages, as described in Ryder et al. (2001).
The relatively compact nature of the
infrared-emitting central region of NGC~7469 allowed sky subtraction to be
performed by sliding the region of interest a total of $38\farcs5$ arcsec along
the slit in between `object' and `sky' exposures. The difference of each
object--sky pair on a given slit position was co-added into groups.
The extraction apertures were defined independently for each grouped
image, in consultation with Fig.~\ref{f:knotpos}. Owing to the still
somewhat coarse spatial resolution of CGS4 when compared with the
typical size and separation of the clusters in the star-forming ring,
it proved impossible to reliably extract the emission from individual
clusters. Instead, the emission from groups of clusters has been
summed, representing the (luminosity-weighted) mean emission from an
ensemble of clusters. This nevertheless allows us to compare the
emission, and thus the SF history, from four distinct
regions along the circumnuclear ring.

\subsubsection{Photometric calibration}

Since the weather
conditions were not photometric during the second night, we made use of
the available {\it HST}/NICMOS F222M ($\sim K$) image to perform the
photometric calibration of the UKIRT spectra. A TinyTim PSF scaled to
best match the bright nuclear emission was subtracted in order to minimize
its contamination to the circumnuclear region. We simulated the sizes and
orientations of the UKIRT extraction apertures (Table~\ref{t:slitpars})
on the F222M image (smoothed to the spatial resolution of
the spectra) and measured the flux densities within
the four slits. In a similar fashion we measured the fluxes of the  
spectroscopic apertures on the rest of the {\it HST} images in order
to obtain the SED of each ring section.
The uncertainties of flux densities were estimated by shifting the
positions of the simulated apertures $\pm\,$1~pixel in the $x$ and $y$
directions over the {\it HST} images, and also by taking into account
flux variations within the simulated apertures when we manually chose
various intensities to scale the TinyTim PSF with the nuclear peak.
Flux densities of the spectroscopic apertures, together with
their uncertainties, are given in Table~\ref{t:aperphot}.
We also obtained integrated photometry through all the {\it HST} filters
(see Table~\ref{t:knotphot}) of the circumnuclear ring region of NGC~7469
for the area enclosed by two
circular apertures of radii 0$\farcs$7 and 2$\farcs$5,
following the analysis of the ring done by G95.

\begin{deluxetable*}{lcccccccc}
\tabletypesize{\scriptsize}
%\rotate
%\tablewidth{0pc}
%\tablenum{}
%\tablecolumns{9}
\tablecaption{\scriptsize Observed flux densities of the spectroscopic apertures and the whole ring}
\tablehead{\colhead{Region} & \colhead{F218W\tablenotemark{(a)}} & \colhead{F330W\tablenotemark{(a)}} & \colhead{F547M\tablenotemark{(a)}} & \colhead{F606W\tablenotemark{(a)}} & \colhead{F814W\tablenotemark{(a)}} & \colhead{F110W\tablenotemark{(a)}} & \colhead{F160W\tablenotemark{(a)}} & \colhead{F222M\tablenotemark{(a)}}}
\startdata
A1 &  16.5$\pm$1.0  &  13.7$\pm$0.5  &  10.5$\pm$0.2  &  11.7$\pm$0.2  &  8.6$\pm$0.3  &  6.9$\pm$0.3  &  5.5$\pm$0.2  &  3.3$\pm$0.1 \\
A2 &  3.4$\pm$0.3  &  4.9$\pm$0.3  &  6.2$\pm$0.2  &  7.5$\pm$0.2  &  6.4$\pm$0.2  &  6.2$\pm$0.2  &  5.4$\pm$0.2  &  3.3$\pm$0.1 \\
A3 &  11.6$\pm$0.8  &  11.8$\pm$0.5  &  10.8$\pm$0.3  &  11.6$\pm$0.3  &  9.61$\pm$0.2  &  7.1$\pm$0.4  &  5.8$\pm$0.2  &  3.4$\pm$0.1 \\
A4 &  4.5$\pm$0.4  &  5.8$\pm$0.5  &  6.8$\pm$0.3  &  8.9$\pm$0.4  &  7.4$\pm$0.2  &  7.1$\pm$0.4  &  6.1$\pm$0.2  &  3.6$\pm$0.1 \\
%A$_{\rm TOT}$ &  32.5$\pm$2.2  &  31.4$\pm$1.5  &  28.1$\pm$0.8  &  32.2$\pm$0.9  &  25.6$\pm$0.7  &  21.2$\pm$1.0  &  17.4$\pm$0.6  &  10.3$\pm$0.3 \\
Ring &  60.6$\pm$2.4  &  62.2$\pm$1.2  &  67.9$\pm$2.0  &  73.6$\pm$2.2  &  63.0$\pm$1.9  &  54.3$\pm$1.6  &  46.5$\pm$1.9  &  27.9$\pm$0.8 \\
\enddata
\tablecomments{\scriptsize (a) Flux densities and uncertainties (10$^{-16}$~erg~s$^{-1}$~cm$^{-2}$~\AA$^{-1}$) for each {\it  HST} filter.}\label{t:aperphot}
\end{deluxetable*}

\subsubsection{Spectroscopic analysis}

\begin{figure}
\epsscale{1.1}
\plotone{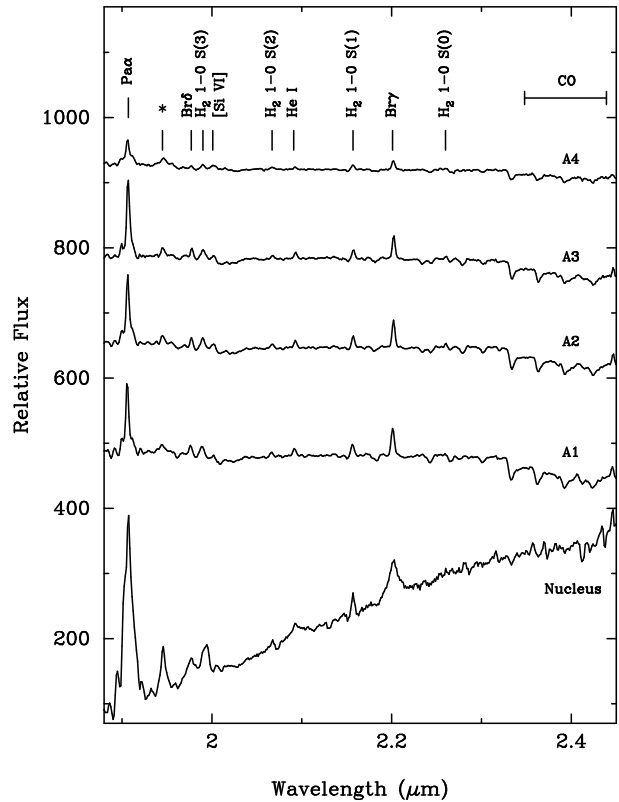}
\vspace{.5cm}
\caption{\footnotesize Spectra of the four UKIRT slit positions along the circumnuclear ring of NGC 7469, as well as through the Seyfert nucleus. The most prominent emission features are labelled, as well as the CO band. The peak at 1.945\,$\mu$m in each spectrum (marked with an asterisk) is spurious, and arises from Br$\delta$ absorption intrinsic to the standard star BS~8738.}\label{f:spec}
\end{figure}

Fig.~\ref{f:spec} shows the spectra of the combined data for each of
the four ring sections (apertures), together with the nuclear spectrum. The
main spectral features present in all spectra are marked at the
top. Line fluxes at $\lambda \le 2\,\mu$m have larger
intrinsic uncertainties due to attempts to correct for atmospheric
transmission, which decreases about 20--40\%
from $1.8-1.96\, \mu$m (these lines are marked with an asterisk in
Table~\ref{t:aperspec}). Although no systematic under/over-correction
of the continuum seems to have been introduced, these values should be
taken with caution.

In order to define a consistent continuum right out into
the region of extensive CO absorption beyond 2.3\,$\mu$m, we fitted a
power-law of the form $F_\lambda \propto \lambda^{\beta}$ to
featureless sections of the spectrum near 2.09 and 2.25~\mic\ (rest
wavelength).
%In practice, the continuum of each slit spectrum, with
%the exception of the nucleus, is quite flat ($\beta \sim -0.2$) as to
%make this step almost unnecessary.
After normalizing each spectrum by this fit, the equivalent widths (EWs) 
and fluxes of the most prominent emission features
were measured, together with the spectroscopic CO$_{\rm sp}$ index, as defined
by Doyon, Joseph \& Wright (1994), and the photometric CO index of 
Frogel et al. (1978): ${\rm CO}_{\rm ph} = \frac{{\rm CO}_{\rm sp} -
0.02}{1.46}$. Line fluxes and EWs, as well as the CO indices are presented 
in Table~\ref{t:aperspec}.

\begin{deluxetable*}{lccccccccc}
\tabletypesize{\scriptsize}
%\rotate
%\tablewidth{0pc}
%\tablenum{}
%\tablecolumns{10}
\tablecaption{\scriptsize Observed line fluxes, EWs and CO indices of UKIRT spectra.}
\tablehead{\colhead{Line\tablenotemark{(a)}} & \colhead{$\lambda_0$\tablenotemark{(b)}} & \colhead{A1} & \colhead{A1} & \colhead{A2} & \colhead{A2} & \colhead{A3} & \colhead{A3} & \colhead{A4} & \colhead{A4} \\
%\cline{3-3} \cline{5-6} \cline{7-8} \cline{9-10}
\colhead{} & \colhead{} & \colhead{Flux\tablenotemark{(c)}} & \colhead{EW\tablenotemark{(d)}} & \colhead{Flux\tablenotemark{(c)}} & \colhead{EW\tablenotemark{(d)}} & \colhead{Flux\tablenotemark{(c)}} & \colhead{EW\tablenotemark{(d)}} & \colhead{Flux\tablenotemark{(c)}} & \colhead{EW\tablenotemark{(d)}}}
\startdata
     Pa$\alpha$* &  1.8751 & 7.97$\pm$0.41 &  23.7$\pm$1.6 & 9.57$\pm$0.35 &  28.6$\pm$1.4 & 12.6$\pm$1.0 &  36.6$\pm$3.0 & 9.94$\pm$0.90 &  25.7$\pm$2.8 \\

     Br$\delta$* &  1.9451 & 1.34$\pm$0.20 &  4.10$\pm$0.70 & 1.29$\pm$0.30 &  4.00$\pm$1.00 & 1.73$\pm$0.28 &  4.90$\pm$0.90 & 1.56$\pm$0.24 &  4.30$\pm$0.70 \\

H$_2$ 1--0 S(3)* &  1.9576 & 1.84$\pm$0.16 &  5.70$\pm$0.50 & 1.91$\pm$0.24 &  6.00$\pm$0.80 & 2.25$\pm$0.23 &  6.70$\pm$0.70 & 2.33$\pm$0.26 &  6.50$\pm$0.70 \\

     $[$SiVI$]$* &  1.9645 & 0.35$\pm$0.05 &  1.10$\pm$0.15 & 0.90$\pm$0.13 &  2.90$\pm$0.55 & 0.60$\pm$0.08 &  1.80$\pm$0.25 & 0.67$\pm$0.14 &  1.90$\pm$0.40 \\

 H$_2$ 1--0 S(2) &  2.0338 & 0.66$\pm$0.07 &  2.00$\pm$0.20 & 0.73$\pm$0.14 &  2.30$\pm$0.50 & 0.56$\pm$0.07 &  1.70$\pm$0.20 & 0.95$\pm$0.15 &  2.70$\pm$0.40 \\

             HeI &  2.0581 & 0.95$\pm$0.10 &  2.90$\pm$0.30 & 0.98$\pm$0.08 &  3.10$\pm$0.30 & 1.04$\pm$0.06 &  3.10$\pm$0.20 & 0.66$\pm$0.14 &  1.80$\pm$0.40 \\

 H$_2$ 1--0 S(1) &  2.1218 & 1.30$\pm$0.09 &  3.90$\pm$0.30 & 1.47$\pm$0.09 &  4.50$\pm$0.30 & 1.38$\pm$0.06 &  4.10$\pm$0.20 & 1.44$\pm$0.13 &  4.10$\pm$0.20 \\

      Br$\gamma$ &  2.1661 & 3.51$\pm$0.13 &  10.7$\pm$0.40 & 3.72$\pm$0.12 &  11.5$\pm$0.40 & 3.60$\pm$0.11 &  10.6$\pm$0.40 & 3.00$\pm$0.11 &  8.40$\pm$0.35 \\

 H$_2$ 1--0 S(0) &  2.2233 & 0.15$\pm$0.05 &  1.45$\pm$0.15 & 0.47$\pm$0.12 &  1.45$\pm$0.40 & 0.39$\pm$0.07 &  1.16$\pm$0.20 & 0.36$\pm$0.06 &  1.02$\pm$0.20 \\

\tableline
%CO (15) & 2.29 & ... & 0.19$\pm$0.05 & ... & 0.21$\pm$0.05 & ... & 0.20$\pm$0.05 & ... & 0.19$\pm$0.05 \\
%CO (17) & 2.32 & ... & 0.23$\pm$0.05 & ... & 0.25$\pm$0.05 & ... & 0.25$\pm$0.05 & ... & 0.22$\pm$0.05 \\
%CO (18) & 2.35 & ... & 0.27$\pm$0.05 & ... & 0.29$\pm$0.05 & ... & 0.27$\pm$0.05 & ... & 0.26$\pm$0.05 \\
CO$_{\rm sp}$  & 2.35 & ... & 0.21$\pm$0.05 & ... & 0.23$\pm$0.05 & ... & 0.23$\pm$0.05 & ... & 0.20$\pm$0.05 \\
CO$_{\rm ph}$  & 2.35 & ... & 0.16$\pm$0.05 & ... & 0.17$\pm$0.05 & ... & 0.17$\pm$0.05 & ... & 0.15$\pm$0.05 \\
\enddata
\tablecomments{\scriptsize (a) Emission lines and CO indices; (b) Rest line/band wavelength ($\mu$m); (c) Flux (10$^{-15}$~erg~s$^{-1}$~cm$^{-2}$) within each aperture; (d) EW in the case of emission lines (\AA); photometric and spectroscopic CO indices in the case of CO bands (see text); * Line fluxes with large uncertainties due to poor atmospheric transmission shortward of $2\, \mu$m (see text).} \label{t:aperspec}
\end{deluxetable*}

The four spectra (see Fig.~\ref{f:spec}) taken at different
positions along the circumnuclear
ring are all rather similar both in terms of line fluxes and their EWs. 
The nuclear spectrum differs from that of the
circumnuclear ring primarily in two respects: first, the extremely red
continuum slope; and second, the broad hydrogen recombination lines.  We
can use these distinguishing characteristics to place an upper limit
on the extent of contamination from the bright nuclear source on the
circumnuclear spectra. A contribution of 2\% or more to the integrated
$K$-band flux from the
nuclear light would be easily noticeable and would have significantly
steepened the flat continua of the ring spectra. The absence of even
such a small contribution means that the $0\farcs6$
slit and UKIRT's active optics facility have done an excellent job at
spatially resolving the circumnuclear starburst ring from the Seyfert
nucleus, despite the latter being only $1\farcs5$ from, but
$\sim\, 3\,$mag brighter than, the region sampled by each slit position.

\section[]{Evolutionary Synthesis Models}\label{s:model}

\subsection{Modelling parameters}\label{s:fitting}

We used Starburst99 (SB99 hereafter; Leitherer et al.
1999; V\'azquez \& Leitherer 2005; with nebular component included) to model
the SF properties of the $1.1\,\mu$m-selected star clusters, the
ring sections, and the ring as a whole.
The main input parameters are: type of SF, initial mass function
(IMF), stellar evolutionary tracks, and metallicity, which was set to solar
(a reasonable assumption for nuclear rings, Mazzuca et al. 2006;
Allard et al. 2007). As the SF processes 
taking place in the nuclear regions of LIRGs and ULIRGs are believed to last
only a few tens of Myr (Kennicutt 1998, Allard et al. 2006, 2007),
we used instantaneous SF bursts. For
the IMF we have used a Salpeter (1955) IMF with lower and upper mass limits
of 0.1~\Msun\ and 100~\Msun, respectively.
Also, because the star clusters located
in the circumnuclear ring of NGC~7469 are expected to be young
($\sim 1-50\,$Myr ---since about 80\% of the $1.1\,\mu$m-selected
clusters are UV-emitters---, Table~\ref{t:ccknots}; see also G95),
we have used the tracks from the Geneva group because they better
reproduce the SEDs of young, ionizing stellar
populations (V\'azquez \& Leitherer 2005).

We have constructed a grid of theoretical SEDs with ages ranging 
from 0.5 to 500\,Myr. The steps of this grid are 0.5~Myr
for the $1-10\,$~Myr age interval, 1~Myr for $10-20\,$Myr, 5~Myr for
$20-100\,$Myr, and 50~Myr for $100-500\,$Myr. The model spectra were
reddened assuming the extinction law of Calzetti et al. 1997 (CAL)
in a simple dust screen configuration which is a good
representation of the dust geometry of individual clusters. This
dust screen model, however, may not be
appropriate for extended regions where more complicated dust geometries are
expected (see discussion by G95).

\subsection{Fitting Method}
We have compared observational and theoretical SEDs through a
$\chi^2$-fitting procedure (Bik et al. 2003):

\begin{equation}\label{e:chi2}
\chi^2_{i,j,k}(t_i,A_{Vj},m_{*,k})=\sum_{N} \frac{(f_{\rm obs}-
f_{\rm model})^2}{\sigma_{\rm obs}^2}
\end{equation}
\noindent where $N$ are the filters or observables in general (photometric
flux densities, flux and/or EW of emission lines, etc.) available for each
cluster, ring section, or entire ring, 
$f_{\rm obs}$ and $f_{\rm model}$ are the observed and model quantities
respectively, and $\sigma_{\rm obs}$ are the weights for the fit. In our case, 
$\sigma_{\rm obs}$ accounts for the photometric calibration 
uncertainties, as well as the uncertainties associated with the theoretical
models. The $\chi^2$ minimization method for fitting SEDs produces
more satisfactory results for determining the ages, extinctions, 
and masses of star clusters detected in galaxies (see, e.g., Maoz et al. 2001; 
Bastian et al. 2005; de
Grijs et al. 2005) than other widely used methods (e.g., color-color diagrams).

When fitting the photometric SEDs (see next sections) we used flux
densities instead of magnitudes to avoid the uncertainties associated with 
transforming fluxes based on the {\it HST}  photometric system to the standard
Johnson-Cousins ($UBVIJHK$) system (see discussion by de Grijs et al. 2005).
To compare the observed flux densities with the outputs of the
evolutionary models, the model spectra (which include the nebular
contribution and the main hydrogen recombination lines) were convolved
with the eight {\it HST} filter bandpasses that take into account both
the filter response and the camera-telescope system throughput.
For the model uncertainties we take 5\% for the optical and NIR, and
10\% for the UV (as assumed by de Grijs et al. 2005).

In the case of a single stellar population,
there is a total of three parameters to be fitted: 
age ($t$), extinction ($A_V$), and stellar mass ($m_{\rm *})$.
To be precise, the stellar mass is
a semi-free parameter because the fitting method simply scales the
theoretical SEDs with the mass until the maximum likelihood
is reached. However, since the mass affects the solution and accuracy of the
fit,  it is necessary to include it in the equation that describes the
method. The expected (minimum) $\chi^2$ value
of the best fit should be equal to the number of degrees of freedom
($\nu = N-3$),  
but in general our fits
provide a higher value of $\chi^2$  (see Table~\ref{t:knotfits}, and next
section). Thus, to estimate the uncertainty of the fitted
parameters, we
define $\nu'=\chi^2_{\rm min}$, where $\chi^2_{\rm min}$ is the
minimum value obtained for a given cluster.
To determine the range of acceptable solutions we take all those solutions
within $\chi^2_{\rm min} \pm \Delta\chi^2_{\rm min}$, with  
$\Delta\chi^2_{\rm min}=\sqrt{2\cdot\nu'}$. This 
would be equivalent to taking the $\pm 1\sigma$ solutions. 
Although this expression is not formally correct, it provides an estimate of
the acceptable solutions when fitting the SEDs.

\begin{deluxetable}{lccccc}
\tabletypesize{\scriptsize}
%\rotate
%\tablewidth{0pc}
%\tablenum{}
%\tablecolumns{5}
\tablecaption{\scriptsize Results of the $\chi^2$-fitting of the $1.1\,\mu {\rm m}$-selected cluster SEDs.}
%\vspace{0.25cm}
\tablehead{\colhead{Cluster} & \colhead{$t$\tablenotemark{(a)}} & \colhead{$A_V$\tablenotemark{(b)}} & \colhead{$m_{\rm *}$\tablenotemark{(c)}} & \colhead{$\chi^2$\tablenotemark{(d)}} & \colhead{$\nu$\tablenotemark{(e)}} \\
\colhead{} & \colhead{(Myr)} & \colhead{(mag)} & \colhead{($\times$ $10^6$~\Msun)} & \colhead{} & \colhead{}}
\startdata
 C1$_{1\#}$ &  13.0$\pm^{  1.0}_{  1.0}$ &  0.00$\pm^{ 0.25}_{ 0.00}$ &   5.9$\pm^{  0.2}_{  0.4}$ &  51.74 & 5 \\
 C2 &  20.0$\pm^{  5.0}_{  4.0}$ &  0.50$\pm^{ 0.25}_{ 0.25}$ &  11.3$\pm^{  1.5}_{  2.4}$ &  11.80 & 5 \\
 C3 &  17.0$\pm^{ 13.0}_{  1.0}$ &  1.25$\pm^{ 0.25}_{ 0.50}$ &  11.5$\pm^{  1.3}_{  1.4}$ &  28.78 & 5 \\
 C4$_{1\#}$ &   2.0$\pm^{  0.5}_{  1.5}$ &  4.50$\pm^{ 0.25}_{ 0.25}$ &  16.1$\pm^{  0.2}_{  3.0}$ &   5.54 & 2 \\
 C5 &  17.0$\pm^{ 18.0}_{  1.0}$ &  1.25$\pm^{ 0.25}_{ 0.50}$ &   9.7$\pm^{  1.3}_{  2.6}$ &  17.08 & 5 \\
 C6$_{1\#}$ &  12.0$\pm^{  1.0}_{  3.5}$ &  1.00$\pm^{ 0.25}_{ 0.25}$ &   3.1$\pm^{  0.6}_{  1.4}$ &  27.98 & 4 \\
 C7$_{1\#}$ &   3.0$\pm^{  0.5}_{  1.0}$ &  1.75$\pm^{ 0.25}_{ 0.25}$ &   4.8$\pm^{  0.2}_{  1.0}$ &   9.92 & 3 \\
 C8$_{1\#}$ &  17.0$\pm^{ 23.0}_{  3.0}$ &  2.75$\pm^{ 0.25}_{ 0.75}$ &  11.9$\pm^{  3.5}_{  7.2}$ &  13.76 & 2 \\
 C9$_{3\#}$ &   8.5$\pm^{  4.5}_{  0.5}$ &  0.50$\pm^{ 0.50}_{ 0.25}$ &   1.1$\pm^{  1.9}_{  0.2}$ &  27.88 & 3 \\
C10 &  20.0$\pm^{ 80.0}_{  5.0}$ &  3.00$\pm^{ 0.25}_{ 0.50}$ &  13.0$\pm^{ 18.2}_{  5.5}$ &   2.75 & 1 \\
C11 &  11.0$\pm^{ 19.0}_{  1.5}$ &  2.50$\pm^{ 1.00}_{ 0.25}$ &   2.5$\pm^{ 11.3}_{  0.6}$ &   2.34 & 1 \\
C13 &  16.0$\pm^{ 14.0}_{  8.0}$ &  1.75$\pm^{ 0.25}_{ 0.75}$ &   4.8$\pm^{  1.8}_{  3.8}$ &   0.46 & 1 \\
C14$_{3\#}$ &  17.0$\pm^{ 23.0}_{ 16.5}$ &  0.75$\pm^{ 1.25}_{ 0.50}$ &   4.9$\pm^{  1.3}_{  2.4}$ &  35.06 & 5 \\
C15$_{1\#}$ &   1.0$\pm^{  1.5}_{  0.5}$ &  2.25$\pm^{ 0.50}_{ 0.25}$ &   3.7$\pm^{  1.2}_{  0.5}$ &  25.36 & 4 \\
C16$_{1\#}$ &  17.0$\pm^{ 13.0}_{  1.0}$ &  1.25$\pm^{ 0.25}_{ 0.50}$ &   4.5$\pm^{  0.4}_{  1.3}$ &   6.30 & 2 \\
C17 &  17.0$\pm^{ 18.0}_{  9.0}$ &  2.00$\pm^{ 0.25}_{ 1.00}$ &   5.4$\pm^{  0.7}_{  4.6}$ &   2.76 & 1 \\
C18 &   8.5$\pm^{  4.5}_{  0.5}$ &  0.75$\pm^{ 0.25}_{ 0.25}$ &   0.8$\pm^{  1.0}_{  0.1}$ &  31.59 & 3 \\
C19 &  12.0$\pm^{  1.0}_{  3.5}$ &  1.50$\pm^{ 0.25}_{ 0.25}$ &   1.6$\pm^{  0.2}_{  0.8}$ &  11.06 & 1 \\
C21 &   2.0$\pm^{ 18.0}_{  1.5}$ &  4.25$\pm^{ 0.25}_{ 2.25}$ &   5.2$\pm^{  2.7}_{  4.2}$ &   5.35 & 1 \\
C22 & 200.0$\pm^{ 50.0}_{194.0}$ &  0.00$\pm^{ 1.25}_{ 0.00}$ &  12.0$\pm^{  0.5}_{ 10.5}$ &   3.26 & 3 \\
C23$_{1\#}$ &   8.5$\pm^{  0.5}_{  0.5}$ &  0.75$\pm^{ 0.25}_{ 0.25}$ &   0.6$\pm^{  0.1}_{  0.1}$ &   1.17 & 1 \\
C24$_{1\#}$ &  18.0$\pm^{ 42.0}_{ 14.0}$ &  0.75$\pm^{ 0.75}_{ 0.50}$ &   2.6$\pm^{  1.8}_{  1.4}$ &   7.06 & 1 \\
C25$_{1\#}$ &   8.5$\pm^{  4.5}_{  0.5}$ &  0.50$\pm^{ 0.25}_{ 0.25}$ &   0.5$\pm^{  0.6}_{  0.1}$ &   1.88 & 1 \\
C28 &   2.0$\pm^{  0.5}_{  1.5}$ &  3.50$\pm^{ 0.25}_{ 0.25}$ &   2.3$\pm^{  0.1}_{  0.4}$ &   7.12 & 2 \\
\enddata
\tablecomments{\scriptsize Results of the fits using SB99 models with an instantaneous SF burst, Geneva tracks, and CAL extinction law; (a) Age; (b) Visual extinction; (c) Stellar mass; (d) Best (lowest) $\chi^2$ value (see Equation~\ref{e:chi2}); (e) Degrees of freedom. The number with a hash symbol next to a cluster label indicates the number of total additional counterparts that have been detected among all of the images (i.e., taken into account all wavelengths; see Table~\ref{t:knotphot} for details).}\label{t:knotfits}
\end{deluxetable}

\section{Small-scale star formation: $1.1\,\mu$m-selected massive star clusters}\label{s:results_indiv}

We have fitted the photometric SEDs of the $1.1\,\mu$m-selected star
clusters using a
single stellar population to derive their ages, extinctions, and masses. 
Since the number of degrees of freedom is $\nu = N-3$, we have only fitted 
those $1.1\,\mu$m-selected star clusters cross-identified in another three
 wavelengths or more; this resulted in a total of 24 fitted clusters
(see Table~\ref{t:knotfits}). Fig.~\ref{f:knotfits}
shows the observational data and the best SED fits of the 10
most luminous clusters sorted by their 
1.1~$\mu$m brightness, as well as four fainter clusters with 3 or more
degrees of freedom. We also show in this figure 
the acceptable range of solutions defined as the area between the
SEDs of the youngest and oldest valid models (within the $1\, \sigma$
uncertainty of the best $\chi^2$ value, as explained in \S~3.2).

\begin{figure*}
\epsscale{1.}
\plotone{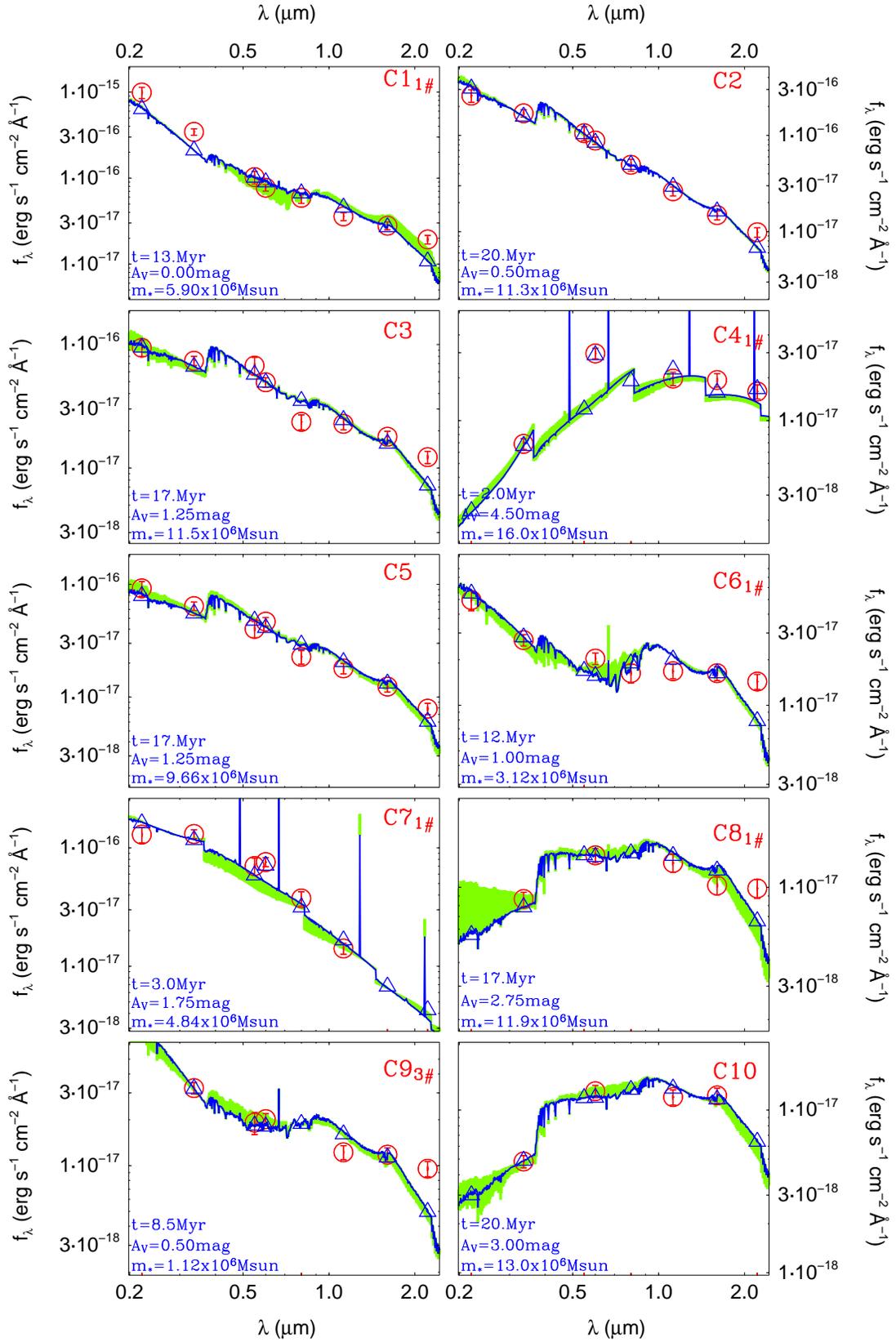}
\caption{\footnotesize Fits to the $1.1\,\mu$m-selected star cluster photometric SEDs. The model SEDs (SB99) use instantaneous star formation, the Geneva tracks, and the CAL extinction law. The ten brightest clusters are shown as well as the remaining four clusters with more than 3 degrees of freedom (see Table~\ref{t:knotfits}) are shown as (red) open circles with their 1 $\sigma$ uncertainties. The (blue) solid line is the best fit of the observational datapoints, and the open, (blue) triangles represent the photometric points resulting from convolving the best fitted SED with the {\it HST}-filter throughputs. The (green) shaded regions represent the SED range of the youngest and oldest acceptable solutions (Table~\ref{t:knotfits}). The emission lines seen in the youngest SEDs are hydrogen recombination lines included in the modelling. As in Table~\ref{t:knotfits}, the number with a hash symbol next to a cluster label indicates the number of total additional counterparts that have been detected among all of the image. [\textit{See the electronic edition of the Journal for a color version of this figure.}]}\label{f:knotfits}
\end{figure*}

%\clearpage

\setcounter{figure}{3}
\begin{figure*}
\epsscale{1.}
\plotone{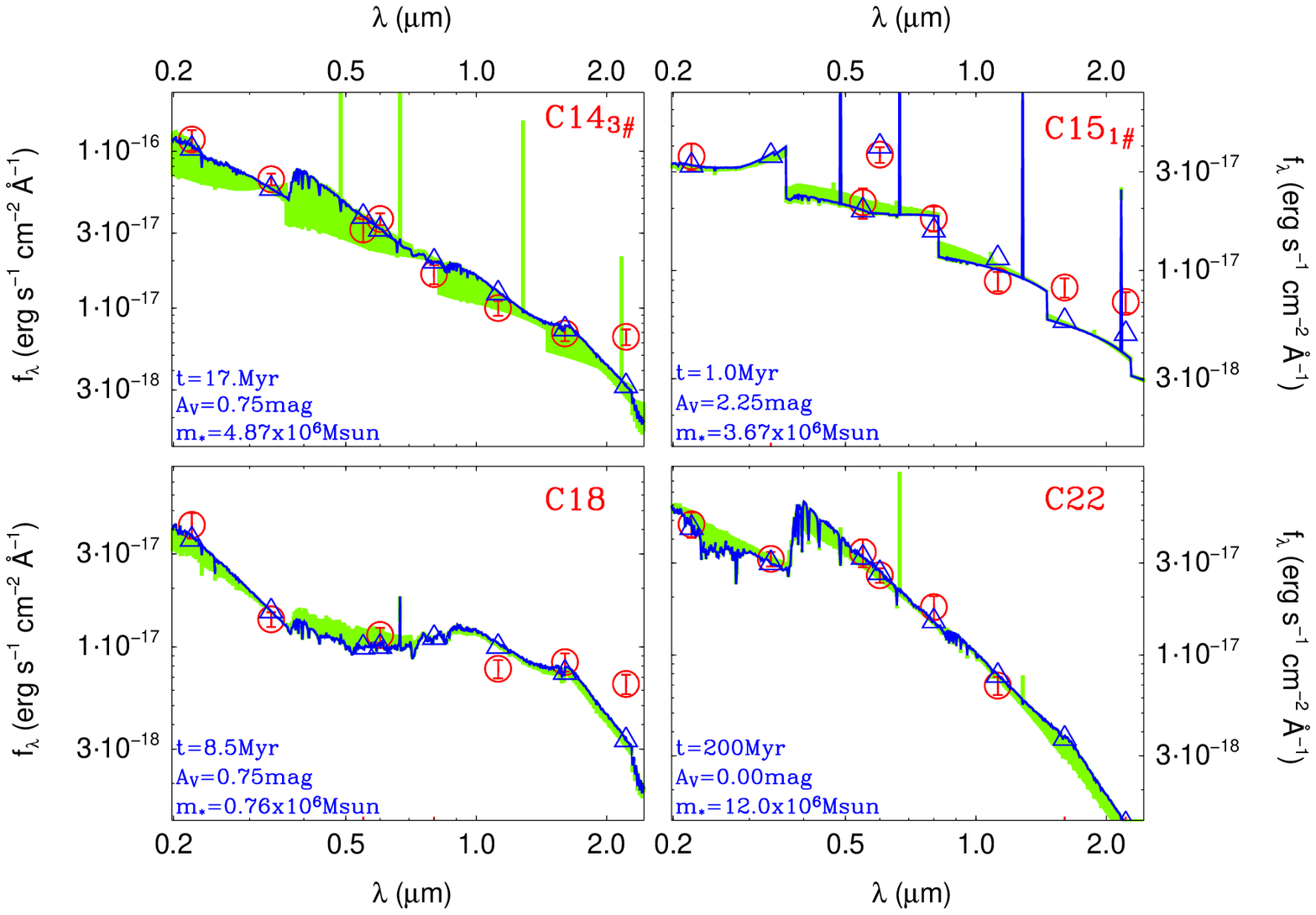}
\caption{\footnotesize Continued. [\textit{See the electronic edition of the Journal for a color version of this figure.}]}
\end{figure*}

The majority (75\%) of  clusters have SEDs best fitted
with ages ranging between 8.5 and 20~Myr. The upper limits of acceptable
ages for these clusters are less constrained because the model SEDs in the
$\sim 10-30\,$Myr range do not vary
as quickly as for younger ($\sim 1-10\,$Myr)
stellar populations. We also note that clusters not detected in the 
UV tend to show large
uncertainties in their fitted parameters. Nearly all of the remaining
($\sim$~21\%) clusters are fitted with a younger, $t=1-3\,$Myr, stellar
population. Only one cluster appears to be relatively old
(C22, 200~Myr), although with a wide range of acceptable ages
(see Table~\ref{t:knotfits}). 
The ages of the $1.1\,\mu$m selected clusters are similar to those of 
clusters detected in other star-forming circumnuclear
rings (Maoz et al. 2001; Alonso-Herrero et al. 2001a), although older
clusters like C22 ($\gtrsim$~100~Myr) have also been detected
in ringed galaxies (Buta et al. 2000).

\begin{figure}
\epsscale{1.2}
\plotone{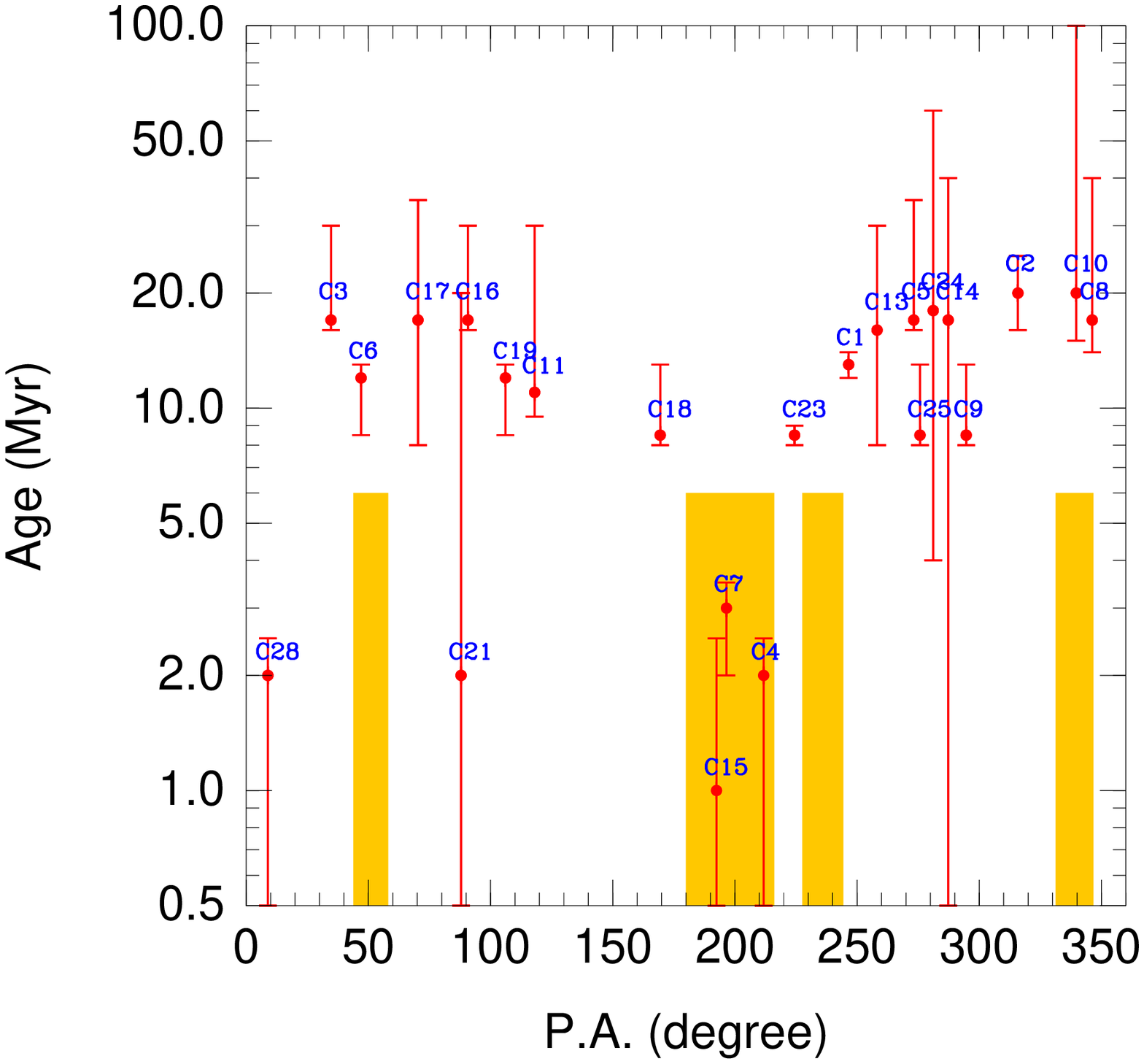}
\caption{\footnotesize Ages of the $1.1\,\mu$m-selected star clusters as a function of their P.A. within the circumnuclear ring. The position angle is measured so that north is 0 degrees and east is 90 degrees. The shaded regions indicate the locations of the MIR/radio (see Fig.~\ref{f:MIR}) peaks tracing the youngest stars ($\lesssim 6-8\,$Myr) in the SF ring. [\textit{See the electronic edition of the Journal for a color version of this figure.}]}\label{f:sequential}
\end{figure}

Fig.~\ref{f:sequential} shows the age
distribution of the clusters as a function of their position angle within
the ring of SF. As well as in age,
the two groups of star clusters also appear to be well differentiated
in terms of their
extinction (see Table~\ref{t:knotfits}). 
The youngest (2.5\,Myr on average) have a mean extinction of
$A_V \sim\, 3\,$mag,
whereas those of intermediate age (on average 14\,Myr) tend to be less
obscured (mean extinction
of $A_V\, \sim\, 1.25\,$mag).
Some clusters (C1, C2, C3, C6, C8, C9, C14, C15, C18 and C19)
present a $2.2\, \mu$m excess ($\sim 20\%$) with respect
to their fitted SEDs. Maoz et al. (2001) found a similar behavior in
clusters detected in NGC~1512 and NGC~5248, and interpreted it as
an additional emission component from dust heated by the star clusters.
The spatial distribution
of the $1.1\,\mu$m-selected star clusters and their relation to the gas
and dust distribution in the ring will be discussed in detail in \S~6.

A large fraction of clusters in the ring of NGC~7469 
has stellar masses  between 0.5 and $10 \times 10 ^6\,$\Msun,
although about one-third appear to be more massive 
($\geq 10 \times 10^6\,$\Msun). 
These values are also in agreement with stellar masses inferred for
SSCs found in other galaxies
%(M51a, Calzetti et al. 2005;
($10^{4}\,$\Msun\,$\lesssim\,$M\,$ \lesssim 10^{6}\,$\Msun\, for M82,
de Grijs, Parmentier \& Lamers 2005)
and especially in LIRGs (from $\sim 10^{4}\,$\Msun\ to few $10^{7}\,$\Msun;
Antennae galaxy, Zhang \& Fall 1999;
NGC~3256 and others, Alonso-Herrero et al. 2000, 2001b, 2002;
Arp~299, Garc\'{\i}a-Mar\'{\i}n et al. 2006; Arp~220, Wilson et al. 2006).
Conversely, other studies suggest that not all
circumnuclear rings show such massive SSCs (Maoz et al. 1996;
Maoz et al. 2001), although these galaxies are not classified as LIRGs.
One possibility is that
some of the $1.1\,\mu$m-selected clusters identified as point sources
in the ring of NGC~7469 are aggregates of less massive clusters, as suggested
by the ACS F330W image (see Fig.~\ref{f:panel}). 

\section{Star formation on the kiloparsec scale: The Circumnuclear Ring}\label{s:results_whole}

\subsection{Stellar Populations}\label{s:param-aper}

The four UKIRT slits (see
Fig.~\ref{f:knotpos}) cover regions with sizes of the order
of $0.2\,{\rm kpc} \times 1.4\,{\rm kpc}$ each, increasing the
likelihood to probe stellar populations with different ages. 
As discussed in \S~4, a large fraction of the
$1.1\,\mu$m-selected star clusters have an average age of 14\,Myr.
Nevertheless, the intense \Brgamma\, hydrogen recombination line emission
measured from the UKIRT spectra has to be produced by a younger
ionizing stellar population.
%Moreover, as we shall discuss in \S~6, the ionizing sources 
%in the ring of NGC~7469  can be traced by the presence of 
%regions with strong MIR emission (e.g., Helou et al. 2004;
%Calzetti et al. 2005; Alonso-Herrero et al. 2006b).
Thus, to fit the photometric SEDs and spectroscopic data (\Brgamma\ flux and
its EW) of the ring sections we used a simple combination of two stellar
populations\footnote{We note that 
a single stellar population cannot simultaneously fit the
photometric SEDs and the spectroscopic data. Specifically,
a 6\,Myr-old population with $A_V\, \sim\, 3\,$mag could
reproduce well the
spectroscopic data, but would not fit the photometric points.}.
Because the ionizing and the MIR emitting regions are spatially coincident
(Soifer et al. 2003 and G95, and discussion in \S~6), the young stellar
population has  
to be at least moderately obscured. Thus, we introduced 
an additional constraint for the extinction of the  
young stellar population: 
$A_V\, \geq\, 3\,$mag, based on the typical extinctions over the
Pa$\alpha$ emitting regions found for local LIRGs (Alonso-Herrero et al.
2006a). From gas and stellar dynamics, G95 inferred a dynamical mass
for the ring of $4.5 \times 10^9\,$\Msun\, whereas Davies et al. (2004)
estimated a dynamical mass of $6.5 \times 10^9\,$\Msun\,
within a radius of 2.5" (with the nucleus contributing at most $\sim 15\%$).
In addition, G95 estimated a total stellar mass of
$2.7 \times 10^9\,$\Msun\, inside the ring. Thus, taking into accout
the uncertainties in all the derived masses, we imposed that the total
stellar mass fitted
for the ring sections (and the whole ring, see below) was not
greater than $3.5 \times 10^9\,$\Msun\, ($\sim 50\%\,$ of $M_{\rm dyn}$).
All the solutions that did not obey this condition were discarded.

\begin{figure*}
\epsscale{1.}
\plotone{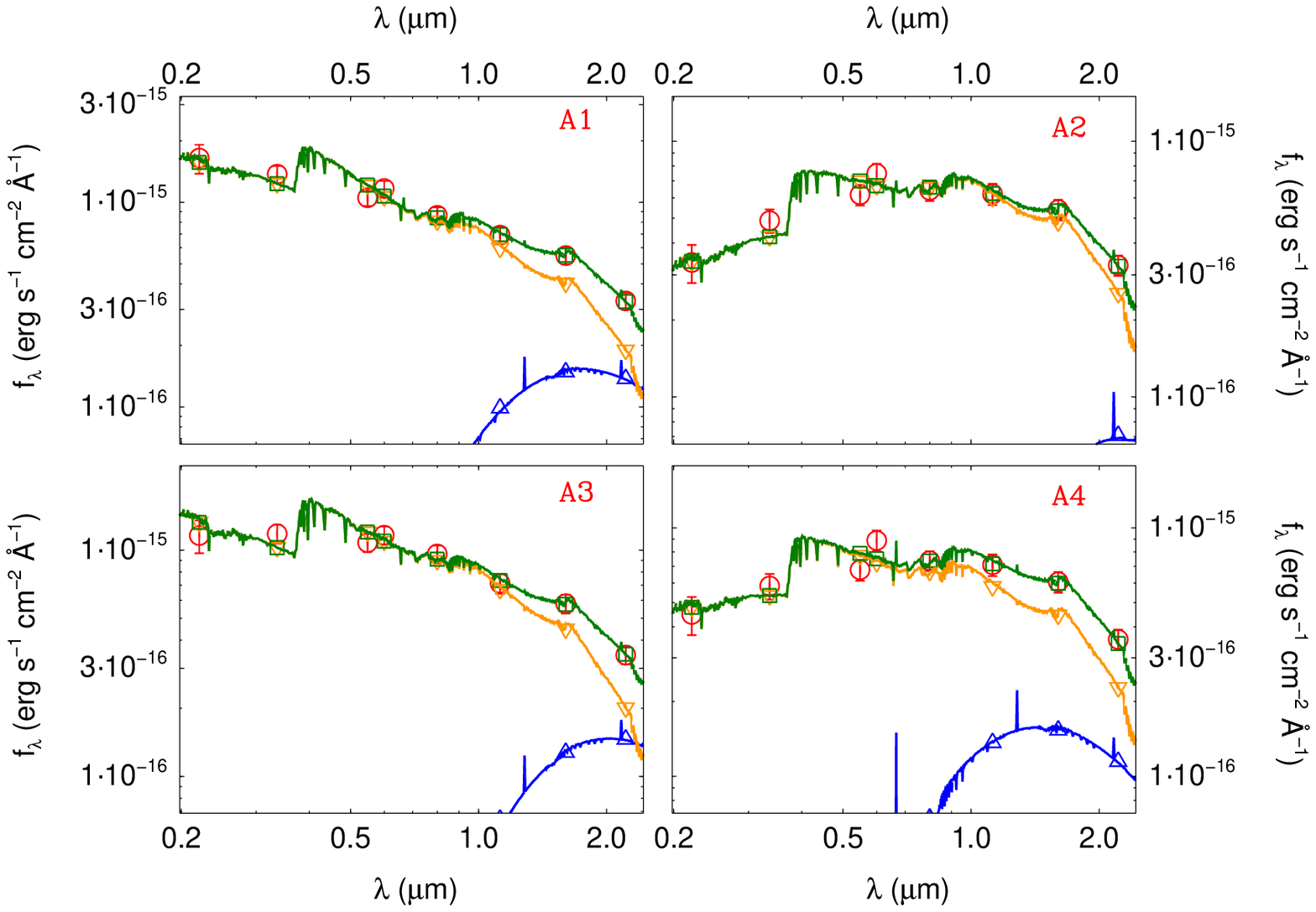}
\vspace{-.5cm}
\caption{\footnotesize Fits to the SEDs (and spectroscopic data, not shown) of the four ring sections (see locations in Fig.~\ref{f:knotpos}). We imposed the extinction of the young stellar population to be $A_{\rm V}\, \geq\, 3\,$mag. The observational datapoints are shown as (red) open circles with their corresponding 1~$\sigma$ uncertainties. The SED of the young stellar population is shown in blue, the intermediate-age population in orange, and the total (sum) fitted SED in green. The (blue) triangles, inverted (orange) triangles, and (green) squares are the model photometric points (SB99 SEDs convolved with {\it HST}-filter throughputs) of the best fit for each population, respectively. [\textit{See the electronic edition of the Journal for a color version of this figure.}]}\label{f:aperfits}
\end{figure*}

The derived masses, ages, and extinctions for the two stellar populations in 
the ring sections are given 
in Table~\ref{t:aperfits} together with the model fits for the 
spectroscopic data (which can be compared with the observations in 
Table~\ref{t:aperphot}) and bolometric luminosity.
Fig.~\ref{f:aperfits} shows the fits to
the photometric SEDs. As can be seen from Table~\ref{t:aperfits} 
the four ring sections have similar fitted properties.
The young stellar population has a
rather restricted age range ($\sim 5-6\,$Myr) and an average extinction of
$A_V \sim 10\,$mag. The fitting of the age is basically driven by the EW
of \Brgamma\, under the assumption that $A_V$(stars) = $A_V$(gas).
However, if the extinction to the gas is higher than that to the stars
(e.g., Calzetti et al. 2000), age would be an upper limit.
The mass in young stars for each aperture is
between $\sim 250$ and $500 \times 10^6\,$\Msun, comparable in most cases
to the mass of the
intermediate-age stellar population (see below). The only ring section with a
mass in young stars significantly smaller than that of the intermediate-age
population is A4 (the eastern part of the ring), 
as suggested from the lack of bright MIR emission there (see \S~6)
and the lower \Brgamma\, emission line flux (when compared to the other
ring sections). The second stellar population in the ring sections 
has an intermediate age ($\sim 17-35\,$Myr)
and is less extincted ($A_V\, \sim\, 1.8\,$mag). Interestingly,
these properties are similar to those of the majority of the
$1.1\,\mu$m-selected clusters (see \S~6).

\begin{deluxetable*}{lccccccccccc}
\tabletypesize{\scriptsize}
%\rotate
%\tablewidth{0pc}
%\tablenum{}
%\tablecolumns{5}
\tablecaption{\scriptsize Results of the $\chi^2$-fitting of the SEDs and spectroscopic data of the ring sections and the whole ring.}
%\vspace{0.25cm}
\tablehead{\colhead{Reg.\tablenotemark{(a)}} & \colhead{} & \colhead{Young pop.\tablenotemark{(b)}} & \colhead{} & \colhead{} & \colhead{Int-age pop.\tablenotemark{(c)}} & \colhead{} & \colhead{} & \colhead{} & \colhead{} & \colhead{} & \colhead{} \\
\cline{2-4} \cline{5-7}
\colhead{} & \colhead{$t$\tablenotemark{(d)}} & \colhead{$A_V$\tablenotemark{(e)}} & \colhead{$m_{\rm *}$\tablenotemark{(f)}} & \colhead{$t$\tablenotemark{(d)}} & \colhead{$A_V$\tablenotemark{(e)}} & \colhead{$m_{\rm *}$\tablenotemark{(f)}} & \colhead{Flux$_{\rm \Brgamma}$\tablenotemark{(g)}} & \colhead{EW$_{\rm \Brgamma}$\tablenotemark{(h)}} & \colhead{log(L$_{\rm bol}$/\Lsun)\tablenotemark{(i)}} & \colhead{$\chi^2$\tablenotemark{(j)}} & \colhead{$\nu$\tablenotemark{(k)}} \\
\colhead{} & \colhead{(Myr)} & \colhead{(mag)} & \colhead{($\times$ $10^6$~\Msun)} & \colhead{(Myr)} & \colhead{(mag)} & \colhead{($\times$ $10^6$~\Msun)} & \colhead{(erg s$^{-1}$ cm$^{-2}$)} & \colhead{(\AA)} & \colhead{} & \colhead{} & \colhead{} }
\startdata
 A1 &   6.0$\pm^{  0.5}_{  0.5}$ &  9.50$\pm^{ 7.00}_{ 3.00}$ &  390$\pm^{ 340}_{ 100}$ &  17.0$\pm^{ 18.0}_{  3.0}$ &  1.50$\pm^{ 0.25}_{ 0.50}$ &   310$\pm^{   40}_{  160}$ &   3.56$\times$10$^{-15}$ & 10.5 & 11.03 & 5.20 & 4 \\
 A2 &   5.5$\pm^{  0.5}_{  0.5}$ & 13.00$\pm^{ 7.00}_{ 7.25}$ &  340$\pm^{ 300}_{ 250}$ &  17.0$\pm^{ 13.0}_{  1.0}$ &  2.50$\pm^{ 0.25}_{ 0.50}$ &   450$\pm^{   40}_{   20}$ &   3.83$\times$10$^{-15}$ & 11.3 & 11.05 & 5.22 & 4 \\
 A3 &   6.0$\pm^{  0.5}_{  0.5}$ & 11.50$\pm^{ 8.50}_{ 6.00}$ &  480$\pm^{ 540}_{ 320}$ &  35.0$\pm^{ 20.0}_{ 20.0}$ &  1.25$\pm^{ 0.50}_{ 0.25}$ &   430$\pm^{  200}_{  190}$ &   3.69$\times$10$^{-15}$ & 10.4 & 11.09 & 5.05 & 4 \\
 A4 &   6.0$\pm^{  0.5}_{  0.5}$ &  7.50$\pm^{12.50}_{ 2.00}$ &  270$\pm^{ 320}_{ 120}$ &  17.0$\pm^{ 13.0}_{  1.0}$ &  2.25$\pm^{ 0.25}_{ 0.50}$ &   390$\pm^{   40}_{   50}$ &   3.03$\times$10$^{-15}$ &  8.5 & 10.93 & 6.54 & 4 \\
% A$_{\rm TOT}$ &   6.0$\pm^{  0.5}_{  0.5}$ &  8.50$\pm^{11.50}_{ 3.00}$ & 1021.5$\pm^{1813.9}_{ 274.7}$ &  17.0$\pm^{ 18.0}_{  3.0}$ &  1.75$\pm^{ 0.25}_{ 0.50}$ &  1021.1$\pm^{  140.0}_{  542.6}$ &   3.74 & 4 \\
 Ring &   5.0$\pm^{  1.0}_{  0.5}$ & 10.50$\pm^{ 7.00}_{ 7.50}$ &  860$\pm^{1230}_{ 380}$ &  14.0$\pm^{  1.0}_{  1.0}$ &  1.75$\pm^{ 0.25}_{ 0.25}$ &  1830$\pm^{   70}_{   70}$ &   2.32$\times$10$^{-14}$ &  9.6 & 11.58 & 9.70 & 5 \\
\enddata
\tablecomments{\scriptsize Results of the fits using SB99 models with an instantaneous SF burst, Geneva tracks, and CAL extinction law. \Brgamma\ flux and its EW have been used as additional data for SED fitting. Two constrains have been imposed to the fits: extinction of the young stellar population have to be $A_{\rm V}\, \geq\, 3\,$mag; total stellar mass must not be greater than $3.5\, \times 10^{9}\,$\Msun\ ($\sim\, 50\%$ of the dynamical mass, see text). (a) Ring sections or whole ring; (b) Parameters of the young population fit; (c) Parameters of the intermediate-age population fit; (d) Age;  (e) Visual extinction; (f) Stellar mass; (g), (h) and (i) Fitted \Brgamma\ flux and EW, and L$_{\rm bol}$, respectively. These values come from the combination of both populations; (j) Best (lowest) $\chi^2$ value (see Equation~\ref{e:chi2}); (k) Degrees of freedom. (g) and (h) fitted values can be compared with those (measured) introduced for the fit (see Table~\ref{t:aperspec}).}\label{t:aperfits}
\end{deluxetable*}

\begin{figure}
\epsscale{1.4}
\plotone{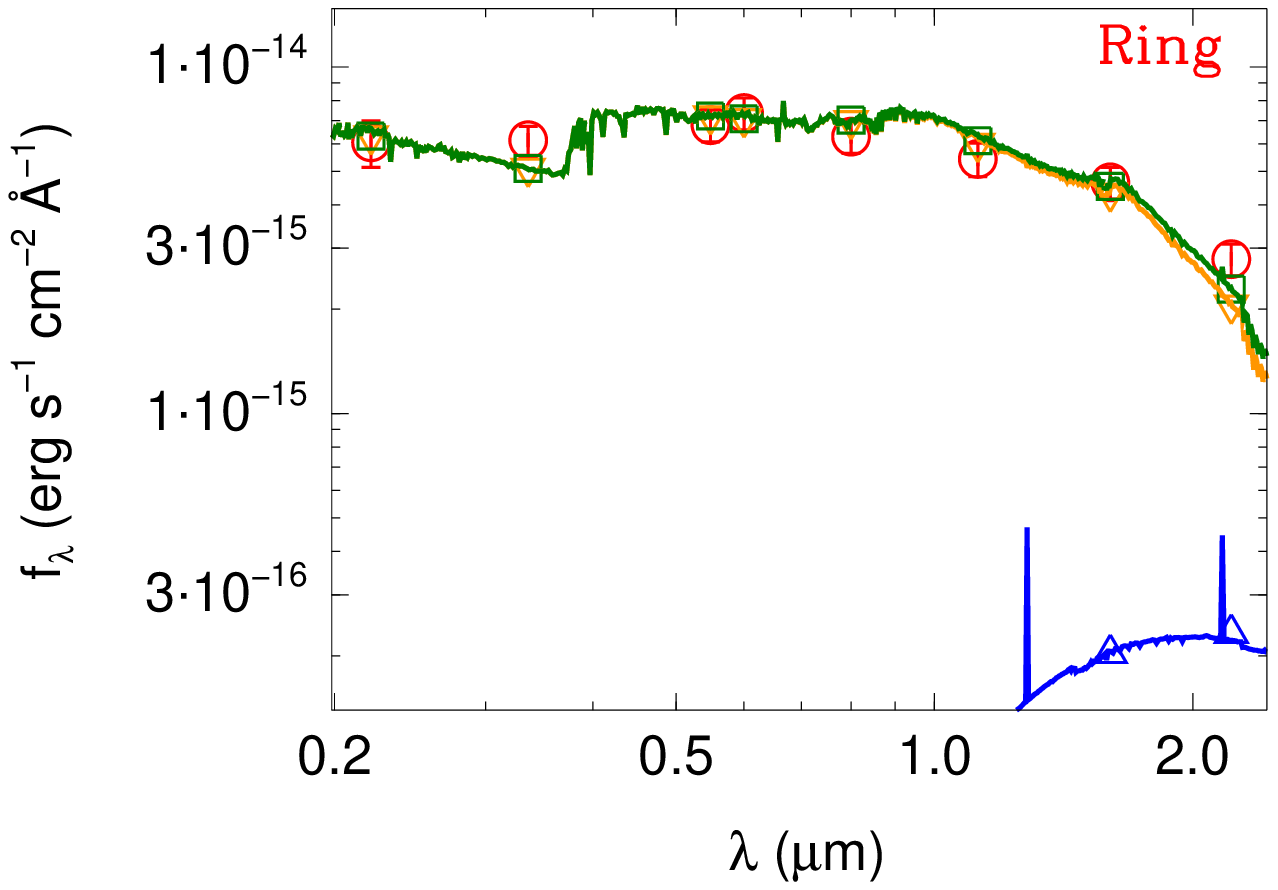}
\vspace{.0cm}
\caption{\footnotesize Fit to the whole ring SED (and spectroscopic data, not shown). We have used the bolometric luminosity of the ring (G95) as an additional datapoint (see \S~5.1 for details). The rest (constraints, symbols, and colors) are the same as in Fig.\ref{f:aperfits}. [\textit{See the electronic edition of the Journal for a color version of this figure.}]}\label{f:ringfit}
\end{figure}

We have also modelled the photometric SED and the spectroscopic
properties of the whole ring, which corresponds to the ring area
used by G95 (see Fig.~\ref{f:knotpos} and Table~\ref{t:aperphot}).
We used the \Brgamma\ flux given by G95
($2.25 \times 10^{-14}\,$erg~s$^{-1}$~cm$^{-2}$), and the EW of \Brgamma\, 
(9.9~\AA) as measured from the mean spectrum of the three (almost)
non-overlapping ring sections (A1, A3 and A4)
covered by UKIRT slits. Besides, we used the bolometric luminosity  
(L$_{\rm IR} \sim\,$L$_{\rm bol} \simeq 3 \times 10^{11}\,$\Lsun,
with a $50\%$ of uncertainty) 
of the ring derived by G95 as an additional datapoint to fit.
The results are presented in Table~\ref{t:aperfits}
and Fig.~\ref{f:ringfit}. 
We find ages of 5~Myr and 14~Myr for the young and intermediate-age
populations, with $A_V \sim 11\,$mag and $A_V \sim 2\,$mag, respectively.
The fitted ages and extinctions are in agreement with those
fitted for the ring sections, not surprisingly, as the three non-overlapping
apertures, which have all similar fitted parameters, account for $\sim 80\%$
of the fitted stellar
mass of the ring (Table~\ref{t:percents}). Thus the ring sections covered
by the UKIRT spectroscopy are representative of the ring as a whole. 
As can be seen from Table~\ref{t:aperfits}, the young
stellar population accounts
for about one-third of the stellar mass of the ring but,
interestingly, is responsible for two-thirds of its bolometric
luminosity.

We can independently estimate the extinction of the ionizing stellar
population of the star-forming ring of NGC~7469. 
A number of authors (e.g., Calzetti et al. 2005; Alonso-Herrero et
al. 2006a) have found a good
correlation between the MIR and the extinction-corrected Pa$\alpha$
luminosities for star forming regions, starbursts, and LIRGs. Thus,
using the $12\,\mu$m vs. Pa$\alpha$ relation shown by Alonso-Herrero et
al. (2006b) and the ring luminosity at $12.5\,\mu$m ($\sim 1\,$Jy)
deduced from Soifer et al. (2003),
we can estimate the number of ionizing photons (corrected for
extinction) for the ring of SF
%(L(\Brgamma)$_{\rm corr}\, \sim\,  4.1\, \times\, 10^{40}\,$erg~s$^{-1}$)
and compare it with the observed \Brgamma\,
flux.
Assuming case B recombination (Hummer \& Storey 1987) and using the CAL
extinction law we derive $A_V \sim 13\,$mag
($A_{{\rm Br}\gamma}\sim 1.25\,$mag). This high value of
the extinction confirms independently (and agrees with)
the existence of a
deeply obscured young population inferred from the
SED analysis of the ring sections and the whole ring (\$~5.1).
%Furthermore, to account for this \Brgamma\ luminosity we would require
%a stellar mass in young stars of $1.2-3.8\, \times\, 10^9\,$\Msun\
%($25-50$\% of the dynamical mass of the ring) for
%a young population of $5-6\,$Myr respectively (using SB99), which
%agrees with our previous findings.

\subsection{Additional Spectroscopic Indicators}\label{s:specindic}

The \HeI\,$2.06\,\mu$m/\Brgamma\, line ratio has been used as an
indicator of the effective temperature of the ionizing stellar population as
well as a means of detecting the presence of massive stars and constraining the
upper mass cutoff of the IMF. Under the assumptions made in this work for
the IMF and metallicity,
we can use this line ratio as an indicator of the age
of the SF processes. For the observed range of the
\HeI\,$2.06\,\mu$m/\Brgamma\, line ratio
($0.22-0.29$ for the spectroscopic apertures,
see Table~\ref{t:aperspec}, and 0.29 for the
ring, see G95) the age of the young
population is approximately $\simeq 5.5-6\,$Myr (using Fig.~3 of
Rigby \& Rieke 2004 for a Salpeter IMF, solar metallicity,
instantaneous
burst and $M_{\rm up}$ = 100~\Msun), in agreement with our previous findings
(see \S~5.1). However, we take these results
with caution as Shields (1993) and Lumsden et al. (2001) have demonstrated 
that this ratio has a complex
dependence with the effective temperature and other H\,{\sc ii} region
parameters, and cannot be used as an accurate indicator of the effective
temperature (Rigby \& Rieke  2004). 

Using SB99 models, the observed photometric CO index (see
Table~\ref{t:aperspec}) is consistent with a stellar population 
with an age of $8-9\,$Myr or $50-100\,$Myr.
Our results would be in agreement with the former interval if we consider
that the two stellar populations fitted in \S~5.1
($\sim\, 6\,$Myr and $\sim 17-35\,$Myr)
would generate an intermediate CO index, 
representative of the mixed populations.
However, as discussed by Origlia \& Oliva (2000) the greatest uncertainty in
deriving ages from CO indices is the stellar evolutionary tracks. This problem
is alleviated if the models use observational NIR stellar libraries (see e.g.,
F$\ddot{o}$rster-Schreiber et al. 2000). 

Summarizing, within the uncertainties of the models, the ages suggested by the
other NIR spectroscopic indicators of the ring sections agree well with the
ages derived from the photometric SED+\Brgamma\, fits discussed in the
previous section, and lead to a consistent result which we will now
place in a wider context.

\section{Discussion}\label{s:discu}

\subsection{Contribution of the $1.1\,\mu$m-selected Star Clusters to
the Integrated Emission}

As we saw in \S~4 and 5.1 most of the $1.1\,\mu$m-selected star clusters
have properties similar to the integrated intermediate-age stellar population
inferred for the ring. However, the flux contribution of the clusters to the
ring sections and the whole ring decreases with wavelength
(Table~\ref{t:percents}). For instance,
the cluster flux contribution to the A1 aperture is 70\% ($40\%$ for the
other ring sections) in the UV and only
$\sim 10\%$ in the NIR. 
This could be in part a spatial resolution effect.
Since the spatial resolution of the UV image is lower than that of
the $1.1\,\mu$m
image, it is possible that an unresolved UV cluster could actually be
an aggregation of
clusters when observed at higher spatial resolution. This is the case
for some UV and NIR clusters which have multiple cross-identifications in the
higher resolution F330W and F606W images (see \S~2.1.1). 
However, the UV and the $2.2\,\mu$m images have similar spatial resolutions,
but the flux contribution of the clusters varies significantly between
these two wavelengths. This suggests that differences in spatial
resolution are not the
main cause. It is more likely that the decreasing flux contribution
(from UV to NIR) of the star clusters to the ring sections could be
due to the existence of an evolved stellar population not spatially
resolved as individual clusters, which would make up a
significant contribution to the NIR emission, but almost none to the
UV continuum emission.

\begin{deluxetable*}{lcccccccccc}
\tabletypesize{\scriptsize}
%\rotate
%\tablewidth{0pc}
%\tablenum{}
%\tablecolumns{5}
\tablecaption{\scriptsize Contribution of the $1.1\,\mu$m-selected clusters to the ring sections and whole ring.}
\tablehead{\colhead{Reg.\tablenotemark{(a)}} & \colhead{Clusters\tablenotemark{(b)}} & \colhead{F218W\tablenotemark{(c)}} & \colhead{F330W\tablenotemark{(c)}} & \colhead{F547M\tablenotemark{(c)}} & \colhead{F606W\tablenotemark{(c)}} & \colhead{F814W\tablenotemark{(c)}} & \colhead{F110W\tablenotemark{(c)}} & \colhead{F160W\tablenotemark{(c)}} & \colhead{F222M\tablenotemark{(c)}} & \colhead{Mass contrib.\tablenotemark{(d)}}}
\startdata
A1 & C1$_{1\#}$,C4$_{1\#}$,C7$_{1\#}$, & & & & & & & & \\
 & C13,C15 & 70\% & 36\% & 18\% & 20\% & 13\% & 12\% & 10\% & 13\% & 5\% \\
A2 & C3,C6$_{1\#}$,C8$_{1\#}$, & & & & & & & & \\
 & C10,C28 & 42\% & 27\% & 10\% & 16\% &  6\% & 11\% & 11\% & 12\% & 5\% \\
A3 & C2,C5,C8$_{1\#}$, & & & & & & & & \\
& C24$_{1\#}$,C25$_{1\#}$ & 44\% & 33\% & 18\% & 20\% &  9\% & 14\% & 12\% & 12\% & 6\% \\
A4 & C3,C6$_{1\#}$,C11, & & & & & & & & \\
& C22,C26,C27 & 45\% & 25\% & 14\% & 13\% &  7\% & 12\% &  9\% & 11\% & 5\% \\
Ring & A1, A3, A4 & 53\% & 51\% & 41\% & 43\% & 40\% & 38\% & 37\% & 36\%   & 80\% \\%53\% & 50\% & 41\% & 43\% & 40\% & 38\% & 37\% & 36\%  & 44\% (1955)\\
\enddata
\tablecomments{\scriptsize (a) Ring sections or whole ring; (b) Clusters (or non-overlapping ring sections, A1, A3 and A4; Fig.~\ref{f:knotpos}) contained in each section (or whole ring, as defined in \S~2.2.1); (c) Fraction of the flux contributed by clusters (or ring sections) to the corresponding section (or the whole ring); (d) Fraction of the stellar mass (young + intermediate-age population) contributed by clusters (or ring sections) to the corresponding section (or whole ring). The number with a hash symbol next to a cluster label indicates the number of total additional counterparts that have been detected among all of the images (i.e., taken into account all wavelengths; see Table~\ref{t:knotphot} for details).}\label{t:percents}
\end{deluxetable*}

In terms of the stellar masses, the $1.1\,\mu$m-selected clusters only
represent a minimal contribution to the total stellar mass ($10\%$ of the
intermediate-age stellar population) within the ring sections. However,
the images with the highest spatial resolutions (ACS F330W and WFPC2 F606W)
reveal a larger number of clusters than those detected at $1.1\,\mu$m 
(Fig.~\ref{f:panel} and Table~\ref{t:ccknots}) whose properties are presumably
similar to those described in \S~4. Given the similar ages and extinctions of
the integrated intermediate-age stellar population detected in the ring and
most of the $1.1\,\mu$m star clusters, it is likely that this ring population
is also made up of less massive clusters ($<1 \times 10^6\,$\Msun), 
not resolved in the NICMOS F110W image. As such, the majority of the
$1.1\,\mu$m-selected clusters would be the most massive (and thus
easier to detect) examples of this intermediate-age stellar population.

\subsection{Gas/dust Distribution and its Relation with the Stellar Populations}\label{s:morph}

\begin{figure}
\epsscale{1.2}
\plotone{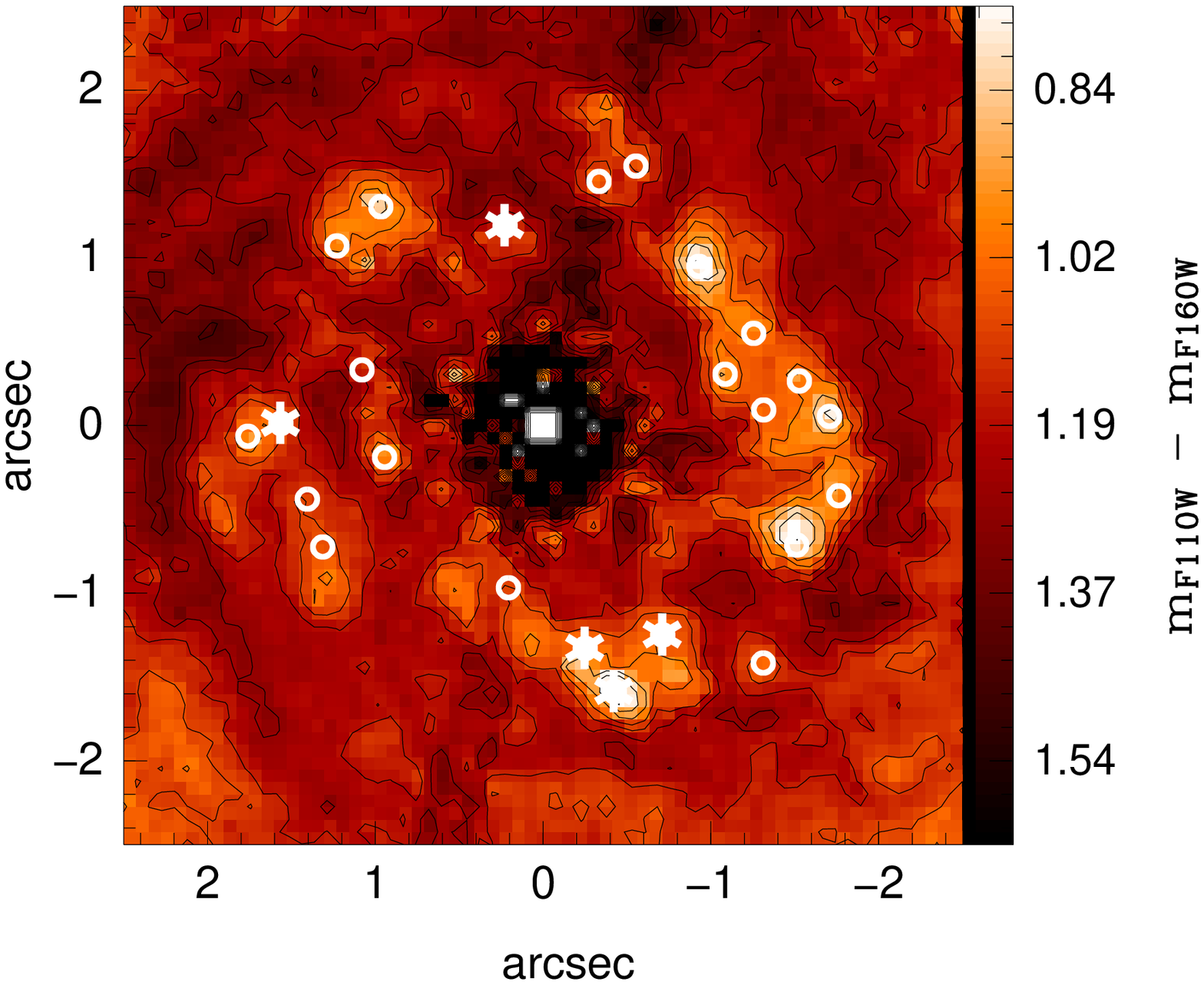}
\vspace{-.5cm}
\caption{\footnotesize $m_{\rm F110W}-m_{\rm F160W}$ color map (equivalent to a $J-H$ color map). Colors redder (darker) than $m_{\rm F110W}-m_{\rm F160W} \sim 1.03$ (typical of an old stellar population, although younger ionizing stellar populations will be bluer, see Alonso-Herrero et al. 2006a) are indicative of highly extincted regions. The young and intermediate-age $1.1\,\mu$m-selected star clusters are marked with star symbols and open circles, respectively. Because of the central region (r $\lesssim$ 0$\farcs$5) is strongly affected by the subtraction of the nucleus, it does not show any real extinction variation. [\textit{See the electronic edition of the Journal for a color version of this figure.}]}\label{f:f110w-f160w}
\end{figure}

The existence of large amounts of obscuring gas and dust in LIRGs
(and ULIRGs), and the major role played by the complexity of their spatial 
structure have been known for a long time (e.g., Sanders et al. 1986; 
Scoville et al. 2000; Bushouse et al. 2002; Alonso-Herrero et al. 2006a,
among many others). This is indeed the case for NGC~7469. The large-scale
{\it HST} imaging (Malkan, Gorjian \& Tam 1998; Scoville et al. 2000) of
NGC~7469 shows that the ring has a dusty clumpy morphology embedded in a
spiral-like structure with prominent dust lanes (Fig.~\ref{f:f110w-f160w}
and Scoville et al. 2000; Martini et al. 2003). A large-scale (several kpc)
stellar bar has been detected by Knapen et al. (2000). In addition, the
CO 2--1 map of Davies et al. (2004) reveals the presence of
molecular gas in the rims of the ring of SF where the dust lanes
are and, more interestingly, forming a nuclear gas bar inside the ring,
crossing the nucleus (see their Fig.~3).

\begin{figure*}
\epsscale{1.}
\plotone{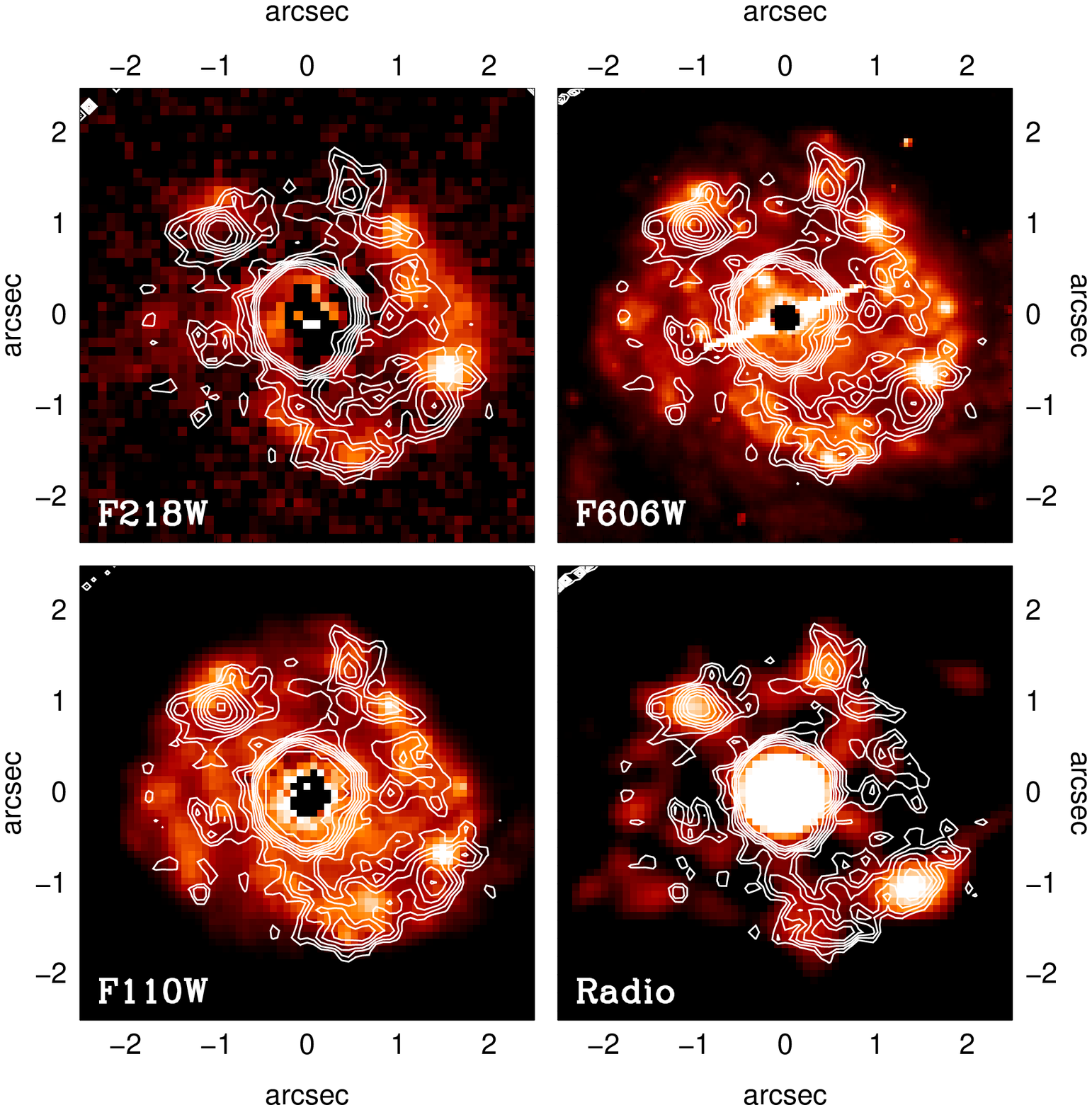}
\caption{\footnotesize In all the images the contours are the MIR $12.5\,\mu$m emission (Soifer et al. 2003) superimposed on the UV F218W (left, top panel), optical F606W (right, top panel), NIR $1.1\,\mu$m (left, bottom panel), and radio 8.4~GHz (right, bottom panel, Colina et al. 2001; obtained in Nov 1999) images. Note the excellent spatial correspondence between the MIR and radio emissions whereas, in general, the locations of brightest NIR, optical, and UV clusters (except those to the south of the nucleus, see \S~6.2) do not correspond with the brightest peaks of MIR emission. [\textit{See the electronic edition of the Journal for a color version of this figure.}]}\label{f:MIR}
\end{figure*}

\begin{figure*}
\epsscale{1.1}
\plotone{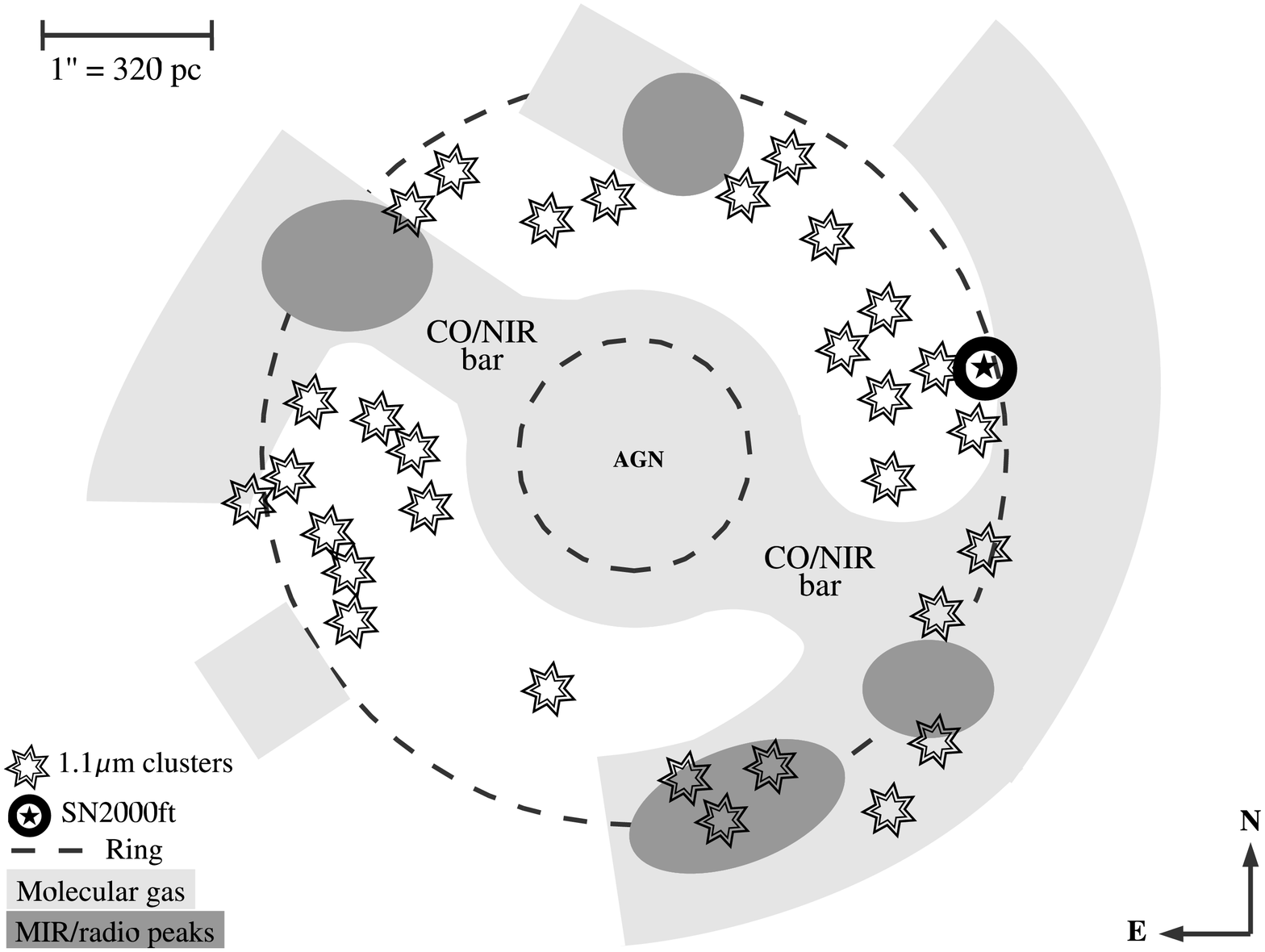}
\vspace{.5cm}
\caption{\footnotesize Sketch of the nuclear region of NGC~7469. The approximate locations of the $1.1 \mu$m-selected star clusters are marked with stars. The places where there is molecular gas (traced by CO J=2-1 emission) are shaded in light gray, whereas the most intense MIR/radio emitting regions are shaded in dark gray. The location of SN2000ft (Colina et al. 2001, 2007) is shown as a black star.}\label{f:sketch}
\end{figure*}

In moderately dusty galaxies, and specially in LIRGs, the 
spatial locations of the young stellar populations can be inferred from
their MIR emission (see, among others, Soifer et al. 
2001; Helou et al. 2004; Calzetti et al. 2005; Alonso-Herrero et
al. 2006b). The $12.5\,\mu$m map of NGC~7469 from Soifer et al. (2003,
see also the
\Brgamma\, map of G95) shows bright MIR peaks in the northeast, south and
southwest regions of the ring (see Fig.~\ref{f:MIR}).
In fact, the ends of the nuclear molecualr gas bar coincide with the position
of these bright MIR peaks (obscured regions of ongoing SF), and these
probably with the location of the Inner Lindblad Resonance(s) of the
large-scale stellar bar (Knapen et al. 1995, Jogee, Scoville \& Kenney 2005).
%At these two locations the gas is expected to be compressed through gravitational instabilities (Jogee, Scoville \& Kenney 2005), creating the conditions necessary for intense SF to take place.
The sketch shown in Fig.~\ref{f:sketch} summarizes the complexity of the
spatial distribution of the stars, dust, and gas in the star-forming
ring of NGC~7469.

As can been seen from Fig.~\ref{f:MIR}, there is a clear
spatial anti-correlation between the positions of the UV-optical-NIR clusters
and the bright MIR emitting regions. The radio (Colina et al. 2001)
and the MIR peaks in the ring of NGC~7469, on the other hand, are spatially
coincindent, as found in other LIRGs (Soifer et al. 2001). 
These displacements between the UV-optical-NIR continuum peaks and
MIR/radio bright emitting regions are real and correspond to distances
of approximately $160\,$pc (0$\farcs$5).

%The spatial anti
%When fitting the properties of the ring sections and the whole ring with 
%instantaneous star formation
% we demonstrated the need for an obscured young ionizing population
%in addition to the intermediate-age ($\sim 10-35\,$Myr)
%stellar population to which
%most of the $1.1\,\mu$m-selected star clusters belong. 

%The radio spectral index of selected regions (see Wilson et al. 1991;
%Colina et al. 2001) is consistent with synchrotron radiation produced
%in SN. Nevertheless, as purely thermal emission (\HII\ regions) is only
%expected in very young populations (P\'erez-Olea \& Colina 1995), this
%indicates that the MIR/radio peaks must be $\gtrsim 4-6\,$Myr, which
%is in agreement with the age of the young, ionizing stellar
%population ($\sim\, 6\,$Myr) obtained from ring section and
%whole ring SED fits.

This lack of spatial coincidence between UV-optical-NIR and MIR continua 
is in agreement with our model of two stellar
populations to fit the integrated properties of the ring (and ring
sections). The UV-optical-NIR continuum (cluster and unresolved emission)
traces mostly the mildly extincted intermediate-age ($\lesssim 100\,$Myr)
population, whereas the MIR and radio peaks reveal the regions where
the most obscured and youngest ($\lesssim 8-10\,$Myr) SF is taking place.
Interestingly, this agrees with the fact that the 
$1.1\,\mu$m-selected clusters, which mostly trace (and are representative of)
the intermediate-age population of the ring (\S~6.1), are indeed located
in low extinction regions (color map, Fig.~\ref{f:f110w-f160w}) that in
general do not coincide with the bright MIR peaks
nor with the molecular gas and dust lanes in the ring
(see Figs.~\ref{f:MIR} and ~\ref{f:sketch}).
That is, most of the clusters have been detected where no \textit{current}
but rather \textit{recent} ($\gtrsim 8-10\,$Myr) SF and low extinction are
expected, which agrees with the parameters found for them
($\sim 14\,$Myr and 1.5 mag of visual extinction).
The only exception are the few young $1.1\,\mu$m star clusters that are
located in the southern part of the ring (where molecular gas,
MIR emisssion, and NIR continuum overlap), which could be the most
massive (luminous), less-extincted examples of the integrated young
stellar population,
as already inferred from the SED fitting of the ring sections (\S~4).
However, due to the small number of young $1.1\,\mu$m-selected clusters,
they might not be statistically representative of the whole young population.

We conclude that in the ring of NGC~7469,
the UV-optical-NIR broad-band continuum
emission reflects the recent SF ($\lesssim 100\,$Myr),
but not necessarily
%\textit{all} of
the ongoing ($\lesssim 8-10\,$Myr) obscured SF. The location and properties
of the youngest population are only revealed by the 
MIR and radio emission, as well as the NIR \Brgamma\, hydrogen recombination
line. In turn, because of the spatially variable and strong obscuration
in the ring of this galaxy, the $1.1\,\mu$m-selected star clusters
mostly trace the intermediate-age, low-extincted stellar population.

\section{Conclusions}\label{s:conclu}

We have presented a multi-wavelength {\it HST} UV-through-NIR
imaging analysis of the
circumnuclear ring of SF of NGC~7469, together with new ground-based
NIR spectroscopy at four locations along the ring. 
These high spatial
resolution data have allowed us to study the SF properties at different
spatial scales, ranging from individual star clusters (tens of parsecs)
to ring sections (hundreds of parsecs) and the ring as a whole
(diameter of $\simeq 1.6\,$kpc).

At the smallest scales we have selected 30 star clusters at $1.1\,\mu$m.
The fitting of their SEDs reveals the presence of two stellar populations.
Most (75\%) of the clusters are of intermediate age ($8-20\,$Myr)
and have low extinction ($A_V \sim 1.25\,$mag). The others are
younger ($\sim 1-3\,$Myr) and are moderately extincted
($A_V \sim 3\,$mag).
Most clusters have stellar masses within the range $1-10 \times 10^6\,$~\Msun,
although more massive clusters or aggregates of clusters  
($10-15 \times 10^6\,$\Msun) are also found.

%Although these clusters are $1.1\,\mu$m selected, they are representative of the entire cluster population detected from the UV to the NIR continuum. 

On the kiloparsec scale (ring sections and whole ring), the
photometric SEDs and spectroscopic indicators 
(in particular \Brgamma\, luminosity and EW) were fitted with two 
distinct stellar populations: a young
($\sim 5-6\,$Myr), highly obscured ($A_V \simeq 7-13\,$mag) population,
and an intermediate-age ($\sim 14-35\,$Myr), less extincted ($A_V
\sim 1.8\,$mag) population. A single stellar population could not
account for all the observed properties.
The young ionizing population would be responsible
for the bright \Brgamma\, line emission and the MIR
and radio continuum emission peaks. This young population
is generally not coincident with the UV-optical-NIR continuum
emitting regions and the majority of the intermediate-age clusters. 
Moreover, the brightest MIR/radio regions coincide with the ends of the
nuclear molecuar gas bar detected in CO.
%and are most probably near the Inner Lindblad Resonance of the large-scale stellar bar of the galaxy.
%(Davies et al. 2004),
%precisely at the locations where SF is expected to be triggered through gravitational instabilities.
This young population of ionizing stars, although it only represents
one-third of the stellar mass of the ring, accounts for a large
fraction (about 2/3) of its IR luminosity. 
The $1.1\,\mu$m-selected clusters, although making up a 
small fraction of the ring stellar mass, are representative of the
integrated intermediate-age
stellar population responsible for the UV-optical-NIR continuum emission.
On the other hand, the younger ($\sim 1-3\,$Myr) clusters could be the less
extincted, massive examples of the young ionizing population traced by
the MIR continuum and \Brgamma\, emission line.

In conclusion, multi-wavelength data (UV-through-radio imaging and NIR
spectroscopy) have allowed us to characterize
the SF properties of both the star clusters and the ring of SF of
NGC~7469. 
In the case of this LIRG, and other dusty galaxies, these types of
observations are critically needed
to discern and study the \textit{current} (traced by MIR/radio
continuum emission and NIR/MIR emission lines) and the \textit{recent}
(detected through UV-optical-NIR continuum) SF taking place
in the dusty environments commonly found in LIRGs and ULIRGs.

\section*{Acknowledgements}

We thank B. T. Soifer and E. Egami for providing us with the MIR
image of NGC~7469. This work has been supported by the
Plan Nacional del Espacio under grant ESP2005-01480.
TDS acknowledges support from the Consejo
Superior de Investigaciones Cient\'{\i}ficas under grant I3P-BPD-2004.
JHK acknowledges support from the Leverhulme Trust in the form of
a Leverhulme Research Fellowship. AAH thanks the Centre for Astrophysics
Research at the University of Hertfordshire for hospitality.
We have made use of observations made with the NASA/ESA Hubble
Space Telescope, obtained from the data archive at the Space Telescope
Science Institute (STScI). The STScI is operated by the Association of
Universities for Research in Astronomy, Inc. under NASA contract NAS 5-26555.
This research has made use of the NASA/IPAC Extragalactic Database
(NED), which is operated by the Jet Propulsion Laboratory, California
Institute of Technology, under contract with the National Aeronautics
and Space Administration, and of NASA's Astrophysics Data System (ADS)
abstract service. The United Kingdom Infrared Telescope is operated by
the Joint Astronomy Centre on behalf of the U.K. Particle Physics and
Astronomy Research Council.\\
\\

%\newpage

%\begin{table*}
%\caption{Spectroscopic Results}\label{t:sliteqw}
%\begin{center}
%\begin{tabular}{lcccc}
%\tableline
%Parameter    &    Slit 1    &     Slit 2    &     Slit 3    &    Slit 4    \\
%\tableline
%$\beta$      & $-0.2\pm0.2$ & $0.1\pm0.2$   & $-0.3\pm0.1$  & $-0.2\pm0.1$ \\ 
%-EW(\Paalpha) (nm) & 
%              $2.80\pm1.00$ & $3.33\pm0.10$ & $3.75\pm0.35$ & $3.44\pm0.9$ \\
%-EW(\Brdelta) (nm) &
%              $0.28\pm0.19$ & $0.39\pm0.15$ & $0.28\pm0.04$ & $0.20\pm0.09$ \\
%-EW(\Brgamma) (nm) &
%              $1.15\pm0.13$ & $1.15\pm0.06$ & $1.13\pm0.05$ & $0.89\pm0.02$ \\
%-EW[H$_{2}$ 1-0 S(1)] (nm) &
%              $0.45\pm0.15$ & $0.44\pm0.03$ & $0.40\pm0.06$ & $0.43\pm0.01$ \\
%H$_{2}$ 1-0 S(0) / H$_{2}$ 1-0 S(1) &
%              $0.27\pm0.16$ & $0.23\pm0.08$ & $0.38\pm0.18$ & $0.35\pm0.05$ \\
%H$_{2}$ 1-0 S(2) / H$_{2}$ 1-0 S(1) &
%              $0.38\pm0.17$ & $0.34\pm0.27$ & $0.33\pm0.20$ & $0.56\pm0.56$ \\
%H$_{2}$ 1-0 S(3) / H$_{2}$ 1-0 S(1) &
%              $0.69\pm0.34$ & $1.16\pm0.42$ & $1.15\pm0.55$ & $1.07\pm0.37$ \\
%-EW([Si$\;${\sc vi}]) (nm) &
%              $0.04\pm0.60$ & $0.15\pm0.60$ & $0.11\pm0.14$ & $0.55\pm0.40$ \\
%-EW(He$\;${\sc i}) (nm) &
%              $0.24\pm0.11$ & $0.27\pm0.01$ & $0.22\pm0.02$ & $0.18\pm0.01$ \\
%CO$_{\rm sp}$     &
%              $0.21\pm0.01$ & $0.25\pm0.02$ & $0.22\pm0.01$ & $0.20\pm0.01$ \\
%\tableline
%\end{tabular}
%\end{center}
%\end{table*}

%\clearpage
%\clearpage
%\clearpage
%\clearpage
%\clearpage
%\clearpage
%\clearpage

\end{document}